\documentclass[a4paper]{article}
\usepackage[a4paper,top=3cm,bottom=2cm,left=3cm,right=3cm,marginparwidth=1.75cm]{geometry}
\usepackage{amsfonts}
\usepackage{bm}
\usepackage{extarrows}
\usepackage{amssymb}
\usepackage{color}
\usepackage[all]{xy}
\usepackage{graphicx}
\usepackage{braket}
\usepackage{amsmath}
\usepackage{appendix}
\usepackage{graphicx}
\usepackage[colorinlistoftodos]{todonotes}
\usepackage{amsthm}
\usepackage{mathrsfs}
\usepackage{amssymb}
\usepackage{accents}
\usepackage{extarrows}
\usepackage{makecell}
\usepackage{authblk}
\usepackage{url}
\usepackage{hyperref}
\usepackage{cite}
\usepackage{eufrak}
\hypersetup{colorlinks,
            citecolor=black,
            linkcolor=black,
            urlcolor=green,
            pdftex}

\usepackage[english]{babel}
\usepackage[utf8x]{inputenc}
\usepackage[T1]{fontenc}

\usepackage[normalem]{ulem}

\title{On the gauge reduction with respect to simplicity constraint in all dimensional loop quantum gravity}
\author[1,2]{Gaoping Long \footnote{201731140005@mail.bnu.edu.cn}\thanks{corresponding author}}
\author[1]{Xiangdong Zhang \footnote{scxdzhang@scut.edu.cn}}

\affil[1]{Department of Physics, South China University of Technology, Guangzhou 510641, China}
\affil[2]{Department of Physics, Beijing Normal University, Beijing 100875, China}

\date{}

\begin{document}

\maketitle

\begin{abstract}
In this paper, we are going to discuss the gauge reduction with respect to the simplicity constraint in both classical
and quantum theory of all dimensional loop quantum gravity. With the gauge reduction
with respect to edge-simplicity constraint being proceeded and the anomalous vertex simplicity constraint being imposed weakly in holonomy-flux phase space, the simplicity reduced holonomy can be established. However, we find that the simplicity reduced holonomy can not capture the degrees of freedom of intrinsic curvature, which leads that it fails to construct a correct scalar constraint operator in all dimensional LQG following the
standard strategy. To tackle this problem, we establish a new type of holonomy corresponding to the simplicity reduced connection, which captures the degrees of freedom of both intrinsic and extrinsic curvature properly. Based on this new type of holonomy, we propose three new strategies  to construct the scalar constraint operators, which serve as valuable candidates to study the dynamics of all dimensional LQG in the future.

\end{abstract}

\section{Introduction}

Loop quantum gravity (LQG) \cite{Ashtekar2012Background}\cite{Han2005FUNDAMENTAL}\cite{thiemann2007modern}\cite{rovelli2007quantum} as a candidate theory of quantum gravity provides a possibility of unifying general relativity (GR) and quantum mechanics. Especially,  the quantum space-time geometry is concealed in some gauge variables and described in a discrete formulation in LQG, and it is an important aspect to derive general relativity (GR) from the foundation of plank-scale quantum geometry.
Indeed, in a broader context, LQG provides a concrete platform for exploring the relation between the continuum classical geometric variables of GR and the discretized geometric quantum data, such as the twistor theory and twisted geometry \cite{Freidel:2010bw}\cite{Freidel:2010aq}.
It has been shown that the correspondence between the field variables of GR and the quantum discrete variables of the geometry of LQG is far beyond the issue of merely taking the continuum limit and semi-classical limit, since the Hamiltonian formulation of GR is governed by a constraint system, and the correspondence could be fully revealed only for the physical degrees of freedom. By this we mean that all the constraints in LQG should be properly imposed to ensure that only all of the physical degrees of freedom are remained. From the opposite direction of this view, the concrete goal of recovering the familiar ADM data from LQG could provide useful instructions in tackling the abstract issues of quantum reductions with respect to the constraints in the theory.

A series of illuminating studies in this direction has been carried out in the case of the $SU(2)$ formulation of $(1+3)$-dimensional LQG. Based on the loop quantization of $SU(2)$ connection formulation of $(1+3)$-dimensional  GR, the kinematic structure of LQG contains the kinematic Hilbert space spanned by the spin-network states and the well-defined $SU(2)$ holonomy-flux operators. Under the actions of holonomy-flux operators, the representations of $SU(2)$-valued holonomies indicate the quanta of the fluxes as the area elements dual to the graph's edges, while the intertwiners relating these representations indicate the intersection angles amongst these fluxes at the vertices. This discretized distribution of the $2$-dimensional spatial area elements with their intersection angles leads to a specific notion of quantum geometry in LQG. The classical constraints--- the scalar, vector and $SU(2)$ Gaussian constraints---can be represented via the holonomy-flux operators for the quantum theory. More explicitly, it has been shown that the imposition of the quantum Gauss constraints on the spin-network states gives rise to a proper quantum gauge reduction, which leads to the reduced state space constituted by the gauge invariant  spin-network states. Remarkably, the gauge invariant spin-network states not only describe the intrinsic spatial geometry built from the polyhedra-cells dual to the network, but also carry precisely the right data to specify the extrinsic curvature of the 3-hypersurface partitioned by these polyhedra \cite{Freidel:2010aq}\cite{Bianchi:2010Polyhedra}\cite{Bianchi:2009ky}. Through this first stage of the gauge reduction with respect to Gaussian constraint, a notion of discrete kinematic ADM data appears in the formulation of Regge geometry, upon which the further reductions with respect to the vector and scalar constraints should be carried out. However, the quantum vector and scalar constraints take much more complicated forms and the quantum algebra between them becomes no longer of first class. At least for now, since the quantum anomaly hinders the standard Dirac procedure from mirroring the classical gauge reduction, the treatment of these loop quantized vector and scalar constraints remains a crucial challenge for LQG tackled by many ongoing projects \cite{Han_2020,Long:2021izw,Zhang:2021qul}.

As we mentioned above, LQG is first established as a quantum theory of GR in four dimensional spacetime. Nevertheless, various classical and quantum gravity theories in higher-dimensional spacetimes (e.g., Kaluza-Klein theory, supergravity and superstring theories) are explored from many different kinds of perspectives. The results of these higher-dimensional theories show remarkable potentials in unifying the gravity and matter fields at the energy scale of quantum gravity. Thus, by extending the framework of loop quantum gravity to higher-dimensional spacetime, one may get a novel approach toward the higher-dimensional ideas of unification, upon the background-independent and non-perturbative construction of the discretized quantum geometry. Pioneered by Bodendorfer, Thiemann and Thurn, the basic framework of loop quantum theory for GR in all dimensions has been developed \cite{Bodendorfer:Ha,Bodendorfer:La,Bodendorfer:Qu,Bodendorfer:SgI}.
The $(1+D)$-dimensional LQG takes the similar framework as
the standard (1+3)-dimensional $SU(2)$ LQG, i.e. the formulation of Yang-Mills gauge theory and the
loop quantization strategy. The key differences between these two theories includes two points. The first one is that the gauge group of $(1+D)$-dimensional LQG is taken as $SO(D + 1)$, while that of the standard (1+3)-dimensional LQG is $SU(2)$. The second key difference is that the $(1+D)$-dimensional LQG contains the simplicity constraint, while the standard (1+3)-dimensional $SU(2)$ LQG does not. Because of the appearance of simplicity constraint,
 the challenge of loop quantum anomaly already exists at the kinematic level before the accounts of the quantum ADM constraints in all dimensional LQG. More explicitly, the all dimensional LQG  is based on the connection formulation of $(1+D)$ dimensional GR in the form of the $SO(D+1)$ Yang-Mills theory, with the phase space coordinatized by the canonical pairs $(A_{aIJ},\pi^{bKL})$, consisting of the spatial $so(D+1)$ valued connection fields $A_{aIJ}$ and the vector fields $\pi^{bKL}$. In this formulation, the theory is governed by the first class constraint system composed by the $SO(D+1)$ Gaussian constraint, the ADM constraints of $(1+D)$-dimensional GR and an additional constraint called the simplicity constraint. The simplicity constraint takes the form $S^{ab}_{IJKL}:=\pi^{a[IJ}\pi^{|b|KL]}$ \cite{Bodendorfer:Ha,Bodendorfer:Qu}, which generates extra gauge symmetries in the $SO(D+1)$ connection phase space. It has been shown that the $SO(D+1)$ connection  phase space correctly reduces to the familiar ADM phase space by proceeding the symplectic reductions with respected to the Gaussian and simplicity constraints. Similar to the $SU(2)$ LQG, the loop quantization of the $SO(D+1)$ connection  formulation leads to the Hilbert space composed by the spin-network states of the $SO(D+1)$ holonomies, with the quantum numbers labeling these states carry the quanta of the flux operators representing the flux of $\pi^{bKL}$ over $(D-1)$-dimensional surfaces. Following the previous study for $SU(2)$ LQG, it is expected to look for the all dimensional Regge ADM data encoded in the $SO(D+1)$ spin-network states, through a gauge reduction procedure with respect to both of the quantum $SO(D+1)$ Gaussian constraint and simplicity constraint.

However, the challenge arises in the gauge reduction procedures with respect to the quantum simplicity constraint--- the quantum algebra among simplicity constraints in all dimensional LQG carries serious quantum anomaly. More explicitly, the commutative Poisson algebra among the classical simplicity constraints becomes the deformed quantum algebra among the quantum simplicity constraint which is not even close \cite{Bodendorfer:2011onthe}. Besides, it has been shown that the ``gauge'' transformations induced by these anomalous quantum simplicity constraints can connect the states which are supposed to be physically distinct in terms of the semiclassical limit. Thus, strong imposition of the anomalous quantum simplicity constraint leads to over-constrained state space which are  not able to capture correct physical degrees of freedom.
Indeed, based on the network discretization, the quantum simplicity constraints in all dimensional LQG are divided into two kinds of local constraints, including the edge-simplicity constraint and the vertex-simplicity constraint. Specifically, the anomaly of quantum algebra only appears for the vertex-simplicity constraint, while the edge-simplicity constraint remains anomaly free in the sense of taking a weakly commutative quantum algebra. To deal with the quantum anomaly of the vertex simplicity constraint, one can focus on the discrete phase space coordinatized by $SO(D+1)$ holonomy-flux variables, in which the Poisson algebras of simplicity constraint are isomorphic to quantum algebras of simplicity constraint and thus the anomaly of vertex-simplicity constraint already exists in the classical holonomy-flux phase space. Previously, based on the so-called generalized twisted geometric parametrization of the edge-simplicity constraint surface, we have proceeded the gauge reduction with respect to  the simplicity constraint in the holonomy-flux phase space \cite{PhysRevD.103.086016}.
 Our result shows that, the discretized classical Gaussian, edge-simplicity constraints and vertex-simplicity
constraint which catches the anomaly of quantum vertex simplicity constraint define a constraint surface
in the holonomy-flux phase space of all dimensional LQG, and the kinematical physical degrees of freedom
 are given by the gauge orbits in
the constraint surface generated by the first class system consisting of discretized Gaussian and edge-simplicity
constraints. We found that, the reduced twisted geometry describes the degrees of freedom of the D-polytopes which partition the D-hypersurface, including the $(D-1)$-faces' areas, the shape of each single D-polytope  and the extrinsic curvature between arbitrary two adjacent  D-polytopes. Finally, the discrete ADM data 
of a D-hypersurface in the form of Regge geometry can be identified as the degrees of freedom of the reduced generalized
twisted geometry space, up to an additional condition called the shape matching condition of $(D-1)$-dimensional faces.  Following this result, this gauge reduction procedures can be realized in quantum theory by imposing the quantum Gaussian and edge-simplicity constraint strongly, and imposing the vertex-simplicity constraint weakly. It leads to the physical kinematic Hilbert space spanned by the spin-network states labelled by simple representations at edges and gauge invariant simple coherent intertwiners at vertices \cite{long2019coherent}.

Nevertheless, the gauge reduction with respect to simplicity constraint has not been accomplished yet, since new issues arise when one consider to construct the gauge invariant operators to describe the kinematic physical observables. Usually, by proceeding the regularization and quantization procedures in LQG, a gauge invariant quantity in connection phase space can be promoted as an operator acting in the physical kinematic Hilbert space, and one could expect that the gauge degrees of freedom are identically eliminated in both classical and quantum case. However, it is not the case for simplicity constraint, since the gauge degrees of freedom with respect to simplicity constraint are eliminated by gauge fixing in classical connection theory, while they are eliminated by taking averaging with respect to gauge transformations in quantum theory.  As we will show in the main part of this article, though the edge-simplicity constraints only transform the pure-gauge components in the holonomy, the gauge reduction with respect to simplicity constraint destroys the structure of holonomy, which leads that the simplicity reduced holonomy can not capture the degrees of freedom of intrinsic curvature. In other words, the simplicity reduced holonomy is not able to inherit the property of connection and thus it can not be used as the building block to regularize the connection. To tackle the problem, we will re-construct the gauge invariant holonomy with respect to the simplicity constraint, which captures the the degrees of freedom of intrinsic and extrinsic curvature properly, by following the geometric interpretation of each component of holonomy given by the twisted geometry parametrization. Besides, we will show that the scalar constraint can be regularized and quantized based on this re-constructed gauge invariant holonomy, with the intrinsic and extrinsic curvatures being catched in the resulting scalar constraint operator properly.

This paper is organized as follows. After our brief review of the classical theory of all dimensional LQG in section \ref{sec2}, we will  introduce the simplicity constraint in both of the connection and holonomy-flux phase space. Especially, we will analyze the gauge degrees of freedom with respect to the simplicity constraint in section \ref{sec2.1} and \ref{sec2.2}. Then, the simplicity reduced holonomy will be constructed and we will also propose a new choice of the gauge (with respect to simplicity constraint) invariant holonomy  in section \ref{2.3}. In section \ref{sec3}, we will turn to consider the gauge reduction with respect to the simplicity constraint in quantum theory of all dimensional LQG. The solution space of quantum simplicity constraint will be introduced firstly, and then the simplicity reduced holonomy operator and a new choice of the gauge (with respect to simplicity constraint) invariant holonomy operator will be considered in our discussions. These operators helps us to consider the construction of the quantum scalar constraint in all
dimensional LQG in section \ref{sec5}. We will first point out the standard strategy is problematic to construct the quantum scalar constraint in section \ref{sec5.1}, and then propose three new strategies for this construction in section \ref{sec5.2}.  Finally, we will finish with a summary and discussion in section \ref{sec6}.
\section{Simplicity constraint in classical theory of (1+D)-dimensional LQG}\label{sec2}
\subsection{Simplicity constraint in connection phase space of (1+D)-dimensional GR}\label{sec2.1}
The connection dynamics of (1+D)-dimensional GR is based on the phase space coordinatized by the canonical field variables $(A_{aIJ}, \pi^{bKL})$ on a spatial D-dimensional manifold $\sigma$, which is equipped with the kinematic constraints---Gauss constraint $\mathcal{G}^{IJ}\approx0$ and simplicity constraint $S^{ab[IJKL]}\approx0$ inducing the gauge transformation of this theory, and the dynamics constraints---vector constraint $C_a\approx0$ and scalar constraint $C\approx0$. More explicitly, the only non-trivial Poisson between the conjugate pair is given by \cite{Bodendorfer:Ha}
\begin{equation}\label{Poisson1}
\{{A}_{aIJ}(x), \pi^{bKL}(y)\}=2\kappa\beta\delta_a^b\delta_{[I}^K\delta_{J]}^{L}\delta^{(D)}(x-y),
\end{equation}
where $\kappa$ is the Newton's gravitational constant, $\beta$ is the Barbero-Immirzi parameter and we used  the notation $a,b,... = 1,2,...,D$ for the spatial tensorial indices and $I,J,... = 1,2,...,D + 1$
for the $so(D+1)$ Lie algebra indices in the definition representation.
 The Gaussian constraint
\begin{equation}
\mathcal{G}^{IJ}:=\partial_a\pi^{aIJ}+2{A}_{aK}^{[I}\pi^{a|K|J]}\approx0,
\end{equation}
simplicity constraint
\begin{equation}
 S^{ab[IJKL]}:=\pi^{a[IJ}\pi^{|b|KL]}\approx0
\end{equation}
combining with the vector constraint and scalar constraint form a first class constraint system in the connection phase space. It has been shown that the symplectic reduction with respect to the Gaussian and simplicity constraints reduces the connection phase space of all dimensional GR to the ADM phase space of dynamics geometry. As one expected, the Gaussian constraint induces the $SO(D+1)$ gauge transformation of the connection ${A}_{aIJ}$ and its momentum $\pi^{bKL}$, while the simplicity constraint restricts the degrees of freedom of $\pi^{aIJ}$ to that of a D-frame $E^{aI}$ to describe the spatial internal geometry and generates some other gauge transformation. The connection variables can be related to the geometric variables on the constraint surface of both Gaussian and simplicity constraint. Specifically, the solution of the simplicity constraint is given by $\pi^{aIJ}=2n^{[I}E^{|a|J]}$ with $E^{aI}$ being the densitized D-frame related to double densitized dual metric by $\tilde{\tilde{q}}^{ab}=E^{aI}E^b_I$ and $n^I$ being a unit internal vector defined by $n_IE^{aI}=0$. Also, one can define the spin connection $\Gamma_{aIJ}$ as
\begin{equation}\label{spingamma}
\Gamma_{aIJ}[\pi]=\frac{2}{D-1}T_{aIJ}+ \frac{D-3}{D-1}\bar{T}_{aIJ}+\Gamma^b_{ac}T^c_{bIJ}
\end{equation}
  which satisfies $\partial_{a}e_{b}^{I}-\Gamma_{ab}^ce_c^I+\Gamma_{a}^{IJ}e_{bJ}=0$ on simplicity constraint surface,  where $T_{aIJ}:=\pi_{bK[I}\partial _a \pi^{bK}_{\ \ J]}$,  $T^c_{bIJ}:= \pi_{bK[I}\pi^{cK}_{\ \ J]}$, $\bar{T}_{aIJ}:=\bar{\eta}_I^K\bar{\eta}_J^LT_{aKL}$, $\bar{\eta}_I^J=\delta_I^J-n_In^J$, $\Gamma_{ab}^c$ is the Levi-Civita connection of $q_{ab}$ and $e_{aI}$ being the D-bein defined by $E^{aI}e_{bI}=\sqrt{q}\delta^a_b$. Based on these conventions, the densitized extrinsic curvature of the spatial manifold $\sigma$ can be given by
\begin{equation}
\tilde{{K}}_a^{\ b}={ K}_{aIJ}\pi^{bIJ}\equiv \frac{1}{\beta}({A}_{aIJ}-\Gamma_{aIJ})\pi^{bIJ}
\end{equation}
 on the constraint surface of both Gaussian and simplicity constraint. Now, it is worth to clarify the gauge transformation induced by simplicity constraint. One can check that $A_{aIJ}$ transforms with respect to simplicity constraint as
\begin{eqnarray}\label{simgauge}
&&\int_{\sigma}d^Dx f_{ab[IJKL]}(x)\{S^{abIJKL}, {A}_{cMN}(y)\}\\\nonumber
&=&2\beta\kappa f_{ac[IJMN]}(y)\pi^{aIJ}(y)= 4\beta\kappa f_{ac[IJMN]}(y)n^{[I}E^{|a|J]}(y).
\end{eqnarray}
 on the simplicity constraint surface. By decomposing the connection ${A}_{aIJ}=2n_{[I}{A}_{|a|J]}+\bar{{A}}_{aIJ}$, it is easy to see that on the simplicity constraint surface, only the component $\bar{{A}}_{a}^{IJ}$ transforms while the component $2n_{[I}{A}_{|a|J]}$ is gauge invariant with respect to simplicity constraint. Similarly, ${K}_{aIJ}
:=\frac{1}{\beta}({A}_{aIJ}-\Gamma_{aIJ})$ can be decomposed as ${K}_{aIJ}=2n_{[I}{K}_{|a|J]}+\bar{{K}}_{aIJ}$.  One can also check that on the simplicity constraint surface, the component $2n_{[I}{K}_{|a|J]}$ is invariant and only $\bar{{K}}_{a}^{IJ}$ transforms under the gauge transformation induced by simplicity constraint. Hence, we see that the simplicity constraint fixes both $\tilde{{K}}_a^{\ b}$ and $q_{ab}$ so that it exactly introduce extra gauge degrees of freedom. In fact, in order to give the gauge invariant variables with respect to simplicity constraint, one can construct the simplicity reduced connection
\begin{equation}
A^S_{aIJ}:=A_{aIJ}-\beta\bar{K}_{aIJ}.
\end{equation}
Then, the symplectic reduction with respect to the simplicity constraint in the connection phase space can be illustrated as
 $$\xymatrix{}
\xymatrix@C=3.5cm{
(A_{aIJ},\pi^{bKL})
\ar[r]^{\textrm{reduction}} &(A^S_{aIJ},\pi^{bKL})|_{S^{abIJKL}=0} },$$
which gives  the gauge invariant variables $(A^S_{aIJ},\pi^{bKL})$ with respect to simplicity constraint on the constraint surface defined by $S^{abIJKL}=0$.

\subsection{Simplicity constraint in discrete phase space of (1+D)-dimensional GR}\label{sec2.2}
Apart from the different gauge group which however is compact and the additional simplicity constraint,
the $SO(D+1)$ connection formulation of (1+D)-dimensional GR is precisely the same as $SU(2)$ connection formulation of (1+3)-dimensional GR, and the quantisation of the $SO(D+1)$ connection formulation is therefore in complete analogy with (1+3)-dimensional $SU(2)$ LQG \cite{Ashtekar2012Background,thiemann2007modern,rovelli2007quantum,RovelliBook2,Han2005FUNDAMENTAL}. By following any standard text on LQG such as \cite{thiemann2007modern,rovelli2007quantum}, the loop quantization of the $SO(D+1)$ connection formulation of (1+D)-dimensional GR leads to a kinematical Hilbert space $\mathcal{H}$ \cite{Bodendorfer:Qu}, which can be regarded as a union of the Hilbert spaces $\mathcal{H}_\gamma=L^2((SO(D+1))^{|E(\gamma)|},d\mu_{\text{Haar}}^{|E(\gamma)|})$ on all possible finite graphs $\gamma$ embedded in $\Sigma$,  where $E(\gamma)$ denotes the set composed by the independent edges of $\gamma$ and $d\mu_{\text{Haar}}^{|E(\gamma)|}$ denotes the product of the Haar measure on $SO(D+1)$. In this sense, on each given $\gamma$ there is a discrete phase space $(T^\ast SO(D+1))^{|E(\gamma)|}$, which is coordinatized by the elementary discrete variables---holonomies and fluxes. The holonomy of $A_{aIJ}$ along an edge $e\in\gamma$ is defined by
 \begin{equation}
h_e[A]:=\mathcal{P}\exp(\int_eA)=1+\sum_{n=1}^{\infty}\int_{0}^1dt_n\int_0^{t_n}dt_{n-1}...\int_0^{t_2} dt_1A(t_1)...A(t_n),
 \end{equation}
 where $A(t):=\frac{1}{2}\dot{e}^aA_{aIJ}\tau^{IJ}$, $\dot{e}^a$ is the tangent vector field of $e$, $\tau^{IJ}$ is a basis of $so(D+1)$ given by $(\tau^{IJ})^{\text{def.}}_{KL}=2\delta^{[I}_{K}\delta^{J]}_{L}$ in definition representation space of $SO(D+1)$, and $\mathcal{P}$ denoting the path-ordered product.
The flux $F^{IJ}_e$ of $\pi^{aIJ}$ through the (D-1)-dimensional face dual to edge $e$ is defined by
\begin{equation}\label{F111}
 F^{IJ}_e:=-\frac{1}{4}\text{tr}\left(\tau^{IJ}\int_{e^\star}\epsilon_{aa_1...a_{D-1}}h(\rho^s_e(\sigma)) \pi^{aKL}(\sigma)\tau_{KL}h(\rho^s_e(\sigma)^{-1})\right),
 \end{equation}
 where $e^\star$ is the (D-1)-face traversed by $e$ in the dual lattice of $\gamma$, $\rho^s(\sigma): [0,1]\rightarrow \Sigma$ is a path connecting the source point $s_e\in e$ to $\sigma\in S_e$ such that $\rho_e^s(\sigma): [0,\frac{1}{2}]\rightarrow e$ and $\rho_e^s(\sigma): [\frac{1}{2}, 1]\rightarrow S_e$. Similarly, we can define the dimensionless flux $X^{IJ}_e$ as
 \begin{equation}
 X^{IJ}_e=-\frac{1}{4\beta a^{D-1}}\text{tr}\left(\tau^{IJ}\int_{e^\star}\epsilon_{aa_1...a_{D-1}}h(\rho^s_e(\sigma)) \pi^{aKL}(\sigma)\tau_{KL}h(\rho^s_e(\sigma)^{-1})\right),
 \end{equation}
 where $a$ is an arbitrary but fixed constant with the dimension of length. Since $SO(D+1)\times so(D+1)\cong T^\ast SO(D+1)$, this new discrete phase space $\times_{e\in \gamma}(SO(D+1)\times so(D+1))_e$, called the phase space of $SO(D+1)$ loop quantum gravity on the fixed graph $\gamma$, is a direct product of $SO(D+1)$ cotangent bundles. Finally, the complete phase space of the theory is given by taking the union over the phase spaces of all possible graphs.
In the discrete phase space associated to $\gamma$, the constraints are expressed by the smeared variables. The discretized Gauss constraints is given by
 \begin{equation}\label{disgauss}
 G_v:=\sum_{b(e)=v}X_e-\sum_{t(e')=v}h_{e'}^{-1}X_{e'}h_{e'}\approx0.
 \end{equation}
The discretized simplicity constraints are separated as two sets. The first one is the edge-simplicity constraint $S^{IJKL}_e\approx0$ which takes the form \cite{Bodendorfer:Qu}\cite{Bodendorfer:SgI}
\begin{equation}
\label{simpconstr}
S_e^{IJKL}\equiv X^{[IJ}_e X^{KL]}_e\approx0, \ \forall e\in \gamma
\end{equation}
and the second one is the vertex-simplicity constraint $S^{IJKL}_{v,e,e'}\approx0$ which is given by \cite{Bodendorfer:Qu}\cite{Bodendorfer:SgI}
\begin{equation}\label{simpconstr2}
\quad S_{v,e,e'}^{IJKL}\equiv X^{[IJ}_e X^{KL]}_{e'}\approx0,\ \forall e,e'\in \gamma, s(e)=s(e')=v.
\end{equation}
The symplectic structure of the discrete phase space can be expressed by the Poisson algebra between the elementary variables $(h_e, X^{IJ}_e)$, which reads
 \begin{eqnarray}
 &&\{h_e, h_{e'}\}=0,\quad\{h_e, X^{IJ}_{e'}\}=\delta_{e,e'}\frac{\kappa}{a^{D-1}} \frac{d}{dt}(e^{t\tau^{IJ}}h_e)|_{t=0}, \\\nonumber
 && \{X^{IJ}_e, X^{KL}_{e'}\}=\delta_{e,e'}\frac{\kappa}{a^{D-1}}(\delta^{IK}X_e^{JL}+\delta^{JL }X^{IK}_e-\delta^{IL}X_e^{JK}-\delta^{JK}X_e^{ IL}).
 \end{eqnarray}
 Based on these Poisson algebras, one can check that the Gaussian constraint generates the $SO(D+1)$ gauge transformation in $SO(D+1)$ Yang-Mills theory, and the edge simplicity constraint induces the transformation
 \begin{equation}\label{esimtrans}
 \{X_e^{[IJ}X_e^{KL]}, h_e\}= 2X_e^{[IJ}\{X_e^{KL]}, h_e\}=-2\kappa\beta X_e^{[IJ}(\tau^{KL]}h_e).
\end{equation}
 Besides, one can evaluate the algebra amongst the discretized Gauss constraints, edge-simplicity constraints and vertex-simplicity constraints. It turns out that $G_v\approx0$ and $S_e\approx0$ form a first class constraint system, with the algebra
\begin{eqnarray}
\label{firstclassalgb}
\{S_e, S_e\}\propto S_e\,,\,\, \{S_e, S_v\}\propto S_e,\,\,\{G_v, G_v\}\propto G_v,\,\,\{G_v, S_e\}\propto S_e,\,\,\{G_v, S_v\}\propto S_v, \quad b(e)=v,
\end{eqnarray}
where the brackets within $G_v\approx0$ are isomorphic to the $so(D+1)$ algebra, and the ones involving $S_e\approx0$ weakly vanish.
Especially, since the commutative momentum Poisson algebra in connection phase space is instead by the non-commutative flux Poisson algebra in the holonomy-flux phase space, the simplicity constraint becomes anomalous at the vertex of the graphs in the holonomy-flux phase space. In other words, the algebras among the vertex-simplicity constraint are the problematic ones, with the open anomalous brackets \cite{Bodendorfer:2011onthe}
\begin{eqnarray}
\label{anomalousalgb}
\{S_{v,e,e'},S_{v,e,e''}\}\propto \emph{anomaly terms}
\end{eqnarray}
where the $ ``\emph{anomaly terms}''$ are not proportional to any of the existing constraints in the phase space.

The anomalous Poisson algebra of the vertex simplicity constraint in discrete phase space destroys the first class constraint system in continuum phase space. Thus, the gauge reduction in discrete phase space can not been a simple copy of the corresponding reduction in continuum phase space. The main obstacle to explore the gauge reduction in discrete phase space is that how to deal with the anomaly of vertex simplicity constraint to reduce correct gauge degrees of freedom. This problem is solve based on the generalized twisted geometric parametrization of the discrete phase space, where the twisted geometry covers the degrees of freedom of the Regge geometries so that it can get back to the connection phase space in some continuum limit \cite{PhysRevD.103.086016}. Let us give a brief introduction of this parametrization as follow.

From now on, let us focus on a graph $\gamma$ whose dual lattice gives a partition of $\sigma$ constituted by D-dimensional polytopes, and the elementary edges in $\gamma$ refers to such kind of edges which only pass through one (D-1)-dimensional face in the dual lattice of $\gamma$. The discrete phase space related to the give graph $\gamma$ is given by $\times_{e\in \gamma}T^\ast SO(D+1)_e$ with $e$ being the elementary edges of $\gamma$. Then, the edge simplicity constraint surface which we are interested in can be given as \cite{PhysRevD.103.086016}
\begin{equation}
\times_{e\in \gamma}T_{\text{s}}^\ast SO(D+1)_e:=\{(h_e,X_e)\in \times_{e\in \gamma}T^\ast SO(D+1)_e|X_{e}^{[IJ}X_{e}^{KL]}=0\}.
\end{equation}
Without loss of generality, we can focus on the edge simplicity constraint surface $T_{\text{s}}^\ast SO(D+1)_e$ related to one single elementary edge $e\in\gamma$.
This space can be parametrized by using the generalized twisted-geometry variables
 \begin{equation}
 (V_e,\tilde{V}_e,\xi_e, \eta_e,\bar{\xi}_e^\mu)\in P_e:=Q^e_{D-1}\times Q^e_{D-1}\times T^*S_e\times SO(D-1)_e,
 \end{equation}
 where $\eta_e\in\mathbb{R}$, $Q^e_{D-1}:=SO(D+1)/(SO(2)\times SO(D-1))$ is the space of unit bi-vectors $V_e$ or $\tilde{V}_e$ with $SO(2)\times SO(D-1)$ is the maximum subgroup fixing the bi-vector $\tau_o:=2\delta_1^{[I}\delta_2^{J]}$ in $SO(D+1)$, $\xi_e\in [-\pi,\pi)$, $e^{\bar{\xi}_e^\mu\bar{\tau}_\mu}:=\bar{u}_e$, and $\bar{\tau}_\mu$ with $\mu\in\{1,...,\frac{(D-1)(D-2)}{2}\}$ is the basis of the Lie algebra of the subgroup $SO(D-1)$ fixing both $\delta_1^{I},\delta_2^{J}$ in $SO(D+1)$. To capture the intrinsic curvature, we specify one pair of the $SO(D+1)$ valued Hopf sections $u_e:=u(V_e)$ and $\tilde{u}_e:=\tilde{u}( \tilde{V}_e)$ which satisfies $V_e=u_e\tau_ou_e^{-1}$ and $\tilde{V}_e=-\tilde{u}_e\tau_o\tilde{u}_e^{-1}$. Then, the parametrization associated with each edge is given by the map
\begin{eqnarray}\label{para}
(V_e,\tilde{V}_e,\xi_e,\eta_e,\bar{\xi}_e^\mu)\mapsto(h_e,X_e)\in T_{\text{s}}^\ast SO(D+1)_e:&& X_e=\frac{1}{2}\eta_e V_e=\frac{1}{2}\eta_eu(V_e)\tau_ou(V_e)^{-1}\\\nonumber
&&h_e=u(V_e)\,e^{\bar{\xi}_e^\mu\bar{\tau}_\mu}e^{\xi_e\tau_o}\,\tilde{u}(\tilde{V}_e)^{-1}.
\end{eqnarray}
 Now we can get back to the discrete phase space of all dimensional LQG on the whole graph $\gamma$, which is just the Cartesian product of the discrete phase space on each single edge of $\gamma$. Then, the twisted geometry parametrization of the discrete phase space on one copy of the edge can be generalized to that of the whole graph $\gamma$ directly. Furthermore, the twisted geometry parameters $(V_e,\tilde{V}_e,\xi_e, \eta_e)$ take the interpretation of the discrete geometry describing the dual lattice of $\gamma$, which can be explained explicitly as follows. We first note that $\frac{1}{2}\eta _e V_e$ and $\frac{1}{2}\eta _e \tilde{V}_e$ represent the area-weighted outward normal bi-vectors of the $(D-1)$-face dual to $e$ in the perspective of source and target points of $e$ respectively, with $\frac{1}{2}\eta _e$ being the dimensionless area of the $(D-1)$-face dual to $e$. Then, the holonomy $h_e=u_e(V_e)\,e^{\bar{\xi}_e^\mu\bar{\tau}_\mu}e^{\xi_e\tau_o}\,\tilde{u}^{-1}_e(\tilde{V}_e)$ takes the interpretation that it rotates the inward normal $-\frac{1}{2}\eta _e\tilde{V}_e$ of the (D-1)-face  dual to $e$ in the perspective of the the target point of $e$, into the outward normal $\frac{1}{2}\eta _e{V}_e$ of the (D-1)-face  dual to $e$ in the perspective of the source point of $e$, wherein $u_e(V_e)$ and $\tilde{u}_e(\tilde{V}_e)$ capture the contribution of intrinsic curvature, and $e^{\xi_e\tau_o}$ captures the contribution of extrinsic curvature to this rotation. Moreover, $\bar{u}_e=e^{\bar{\xi}_e^\mu\bar{\tau}_\mu}$ represents some redundant degrees of freedom for reconstructing the discrete geometry, and it is also the gauge degrees of freedom with respect to edge-simplicity constraint exactly.
Then, beginning with the twisted geometry parameter space $P_\gamma=\times_{e\in\gamma}P_e, P_e:=Q^e_{D-1}\times Q^e_{D-1}\times T_e^*S\times SO(D-1)_e$ related to $\gamma$, the gauge reduction with respect to the kinematic constraints---Gauss constraint and simplicity constraints---can be done by the guiding of their geometrical meaning in Regge geometry in the subset with $\eta_e\neq 0$. Up to a double-covering symmetry, we firstly reduce the $SO(D-1)_e$ fibers for each edge $e$ to get the phase space $\check{P}_\gamma:=\times_{e\in \gamma}\check{P}_e$ with $\check{P}_e:=Q^e_{D_1}\times Q^e_{D-1}\times T^*S_e^1$. Then, the discretized Gauss constraint \eqref{disgauss} can be imposed to give the reduced phase space
 \begin{equation}
\check{H}_\gamma:=\check{P}_\gamma/\!/SO(D+1)^{V(\gamma)}=\left(\times_{e\in\gamma} T^\ast S_e^1\right)\times \left(\times_{v\in\gamma} \mathfrak{P}_{\vec{\eta}_v}\right)
\end{equation}
 with $V(\gamma)$ being the number of the vertices in $\gamma$ and
 \begin{equation}
 \mathfrak{P}_{\vec{\eta}_v}:=\{(V_{e_1}^{IJ},...,V_{e_{n_v}}^{IJ})\in \times_{e\in\{e_v\}}Q_{D-1}^{e}| G_{v}=0 \}/SO(D+1),
 \end{equation}
where we re-oriented the edges linked to $v$ to be out-going at $v$ without loss of generality, $\{e_v\}$ represents the set of edges beginning at $v$ with $n_v$ being the number of elements in $\{e_v\}$, and $G_{v}=\sum_{\{e_v\}}\eta_{e_v}V_{e_v}^{IJ}$ here. Further, we solve the vertex simplicity constraint equation \eqref{simpconstr} in the reduced phase space $\check{H}_\gamma$ and get the final generalized twisted geometric space $\check{H}^{s}_\gamma=\left(\times_{e\in\gamma} T^\ast S_e^1\right)\times \left(\times_{v\in\gamma} \mathfrak{P}^{s}_{\vec{\eta}_v}\right)$ with $\mathfrak{P}^{s}_{\vec{\eta}_v}:=\mathfrak{P}_{\vec{\eta}_v}|_{S_v=0}$. It has been shown that the generalized twisted geometry in the space $\check{H}^{s}_\gamma$ is consistent with the Regge geometry on the spatial D-manifold $\sigma$ if the shape match condition in the D-polytopes' gluing process is considered, which means the gauge reduction scheme in the parametrization space captures the correct physical degrees of freedom of all dimensional LQG in kinematical level. Thus, based on this twisted geometry parametrization, one can conclude that, in order to get correct kinematical physical degrees of freedom, the anomalous vertex should be treated as a second class constraint while the Gauss constraint and edge simplicity constraint are treated as first class constraint in discrete and quantum theory of all dimensional LQG. The reduction procedures can be roughly illustrated as follows \cite{PhysRevD.103.086016}.
\begin{equation}\label{flowchart111}
\xymatrix{}
\xymatrix@C=1.5cm{
\times_{e\in\gamma}T^\ast SO(D+1)_e
\ar[r]_{\ \  \  \ \text{(i)}}&
\times_{e\in\gamma}\check{P}_e\ar[r]_{\text{(ii)}} & \check{H}_\gamma^{}
\ar[r]_{\text{(iii)}}&
\check{H}_\gamma^{s} },
\end{equation}
where  the symplectic reductions with respect to edge simplicity constraint and Gaussian constraint are proceeded in step (i) and (ii) respectively, and in step (iii) the vertex  simplicity constraint equation is solved.

The reduction of holonomy-flux phase based on twisted geometry parametrization can be related to the reduction of the connection phase space by taking the continuum limit \cite{PhysRevD.103.086016}. Observe that the choice for the Hopf sections is clearly non-unique, and the twisted geometric parametrization is given under one fixed choice of $\{u_e,\tilde{u}_e\}$ for every edge $e$, under which the Levi-Civita holonomy $h^{\Gamma}_{e}$ can be expressed in the form
\begin{equation}\label{hgamma}
h^{\Gamma}_{e}( {V}_{e'},\tilde{V}_{e'})\equiv u_e\,\, (e^{\bar{\zeta}_{e}^{\mu}\, \bar\tau_\mu}\, e^{\zeta_{e}\,\tau_o})\,\,\tilde{u}_e^{-1},\quad e'\in\{\{E(b(e))\},\{E(t(e))\}\},
\end{equation}
where $\{E(b(e))\}$ and $\{E(t(e))\}$ are the collections of edges linked to the beginning point $b(e)$ of $e$ and  the target point $t(e)$ of $e$ respectively, $ e^{\bar{\zeta}^\mu \bar\tau_\mu}$ takes value in the subgroup $SO(D-1)\subset SO(D+1)$ preserving both $\delta^I_1$ and $\delta^I_2$. Note that the bi-vector functions $\zeta_{e}$ and $\bar{\zeta}_{e}^{\mu}$ are well-defined via the given $h^{\Gamma}_{e}$ and the chosen Hopf sections. Then, let us take the continuum limit that makes the coordinate length of each edges of $\gamma$ tends to $0$, we get
 \begin{equation}\label{hAcon}
 h_e=u_ee^{\xi_e\tau_o}e^{\bar{\xi}_e^\mu\bar{\tau}_\mu}\tilde{u}_e^{-1}\simeq \mathbb{I}+A_e,
 \end{equation}
\begin{equation}
X^{e'}\simeq\pi^{e'}
\end{equation}
and
 \begin{equation}\label{hgamma}
 h^{\Gamma}_e=u_e\,\, (e^{\bar{\zeta}_{e}^{\mu}\, \bar\tau_\mu}\, e^{\zeta_{e}\,\tau_o})\tilde{u}_e^{-1}\simeq \mathbb{I}+\Gamma_e
 \end{equation}
Furthermore, let us factor out $h^{\Gamma}_{e}$ from $h_e$ through the expressions
\begin{eqnarray}
\label{decomp3}
h_e= h^{\Gamma}_{e}\,\, \left(e^{-\bar{\zeta}_e^\mu \tilde{u}_{e}\!\bar\tau_\mu\!\tilde{u}^{-1}_{e}}\,e^{\bar{\xi}_e^\mu \tilde{u}_{e}\!\bar\tau_\mu\!\tilde{u}^{-1}_{e}}\, e^{-(\xi_e- \zeta_e) \tilde{V}_e}\right) =\left(e^{\bar{\xi}_e^\mu {u}_{e}\!\bar\tau_\mu\!{u}^{-1}_{e}}\,e^{ -\bar{\zeta}_e^\mu {u}_{e}\!\bar\tau_\mu\!{u}^{-1}_{e}}\, e^{(\xi_e- \zeta_e) {V}_e}\right)\,\,h^{\Gamma}_{e}.
\end{eqnarray}
Recall the splitting
\begin{equation}\label{split2}
A_{a}^{IJ}=\Gamma_a^{IJ}(\pi)+\beta K_a^{IJ}
\end{equation}
with $\Gamma_a^{IJ}(\pi)$ being a function of $\pi^{bKL}$ satisfying $\Gamma_a^{IJ}(\pi)=\Gamma_a^{IJ}(e)$ on simplicity constraint surface, and notice Eqs.\eqref{hAcon} and \eqref{hgamma}, we have the continuum limit
\begin{equation}
K_e  \simeq \frac{1}{\beta}u_e (\xi^o_e\tau_o+\check{\xi}_e^\mu\bar{\tau}_\mu)u_e^{-1},
\end{equation}
where $\xi^o_e:=\xi_e-\zeta_e$ and $e^{\check{\xi}^\mu_e\bar{\tau}_\mu}:=e^{-\bar{\zeta }^\mu_e\bar{\tau}_\mu}e^{\bar{\xi}^\mu_e\bar{\tau}_\mu}$.
 Denote $K^{\perp}_e:=\frac{1}{\beta}u_e (\xi^o_e\tau_o)u_e^{-1}$ and $K^{/\!/}_e:=\frac{1}{\beta}u_e (\check{\xi}_e^\mu\bar{\tau}_\mu)u_e^{-1}$, we can clearly see that despite of the anomaly in the vertex-simplicity constraints, our reduction procedure correctly removes the component $K^{/\!/}_e$, while preserves the component $K^{\perp}_e$ that contributes to the extrinsic curvature as expressed in the same form as in the classical connection formulation. In other words, we have
\begin{equation}\label{Kpi}
\text{tr}(K_e\pi^{e'})\simeq\frac{1}{\beta} \text{tr}( u_e (\xi_e^o\tau_o+\check{\xi}_e^\mu\bar{\tau}_\mu)u_e^{-1}X^{e'}) =\frac{1}{\beta}\text{tr}( u_e (\xi_e^o\tau_o)u_e^{-1}X^{e'})=\text{tr}(K^{\perp}_eX^{e'}), \  \  b(e)=b(e')
\end{equation}
in continuum limit.
Indeed, on the constraint surface of both edge-simplicity and vertex-simplicity constraints, the component $K^{/\!/}_e$ has no projection on the bivector $\pi^{e'}\simeq X^{e'}=N_{e'}V_{e'}$ satisfying $V_e^{[IJ}V_{e'}^{KL]}=0$, thus it provides no contribution to the extrinsic curvature as it showed in above Eq.\eqref{Kpi}.
Then, one can conclude that
  the degrees of freedom of $\check{\xi}_e^\mu$ are consistent with the gauge component $\bar{K}_{aIJ}$ for non-anomalous simplicity constraint in continuum phase space in continuum limit, so that the components $\check{\xi}_e^\mu$ are regarded as pure gauge (with respect to simplicity constraint) component in discrete phase space, which can be illustrated as \cite{PhysRevD.103.086016}
 \begin{equation}\label{gaugecorresp}
\bar{K}_{aIJ}\overset{\textrm{correspondence of gauge degrees of freedom}}{\underset{\textrm{ in continuum limit} }{\leftarrow--------------\rightarrow}}\check{\xi}_e^\mu.
\end{equation}

\subsection{Classical gauge reduction with respect to simplicity constraint}\label{2.3}
 To construct the gauge invariant variables with respect to edge-simplicity constraint in the holonomy-flux phase space, one need to reduce the holonomy and flux variables respectively. Let us focus on the constraint surface defined by edge-simplicity constraint in the phase space $T^\ast SO(D+1)$ associated to one single elementary edge $e$ of $\gamma$. Based on the twisted geometry parametrization, we note that the gauge transformation induced by edge-simplicity constraint on the edge-simplicity constraint surface is given by
 \begin{eqnarray}\label{edgesimtrans2}
 \{X_e^{[IJ}X_e^{KL]}, h_e\}&=& 2X_e^{[IJ}\{X_e^{KL]}, h_e\}
 \propto \eta_eV_e^{[IJ}(\tau^{KL]}u_e\,e^{\bar{\xi}_e^\mu\bar{\tau}_\mu}e^{\xi_e\tau_o}\,\tilde{u}_e^{-1}) \\\nonumber&=&\eta_e(u_e\,(\bar{\tau}_e^{IJKL}e^{\bar{\xi}_e^\mu\bar{\tau}_\mu}) e^{\xi_e\tau_o}\,\tilde{u}_e^{-1})
\end{eqnarray}
and
\begin{equation}
\{X_e^{[IJ}X_e^{KL]}, X^{MN}_e\}=0,
\end{equation}
 where we defined $\bar{\tau}_e^{IJKL}:=V_e^ {[IJ}(u_e^{-1}\tau^{KL]}u_e)\in so(D-1)$. It easy to see that the edge simplicity constraint induce the transformation of the component $e^{\bar{\xi}_e^\mu\bar{\tau}_\mu}\in SO(D-1)$ in the parametrization of $h_e$, and the flux is gauge invariant with respect to edge-simplicity constraint on the constraint surface defined by edge-simplicity constraint. Thus, we only need to focus on the reduction of holonomy.
Let us introduce the averaging operation $\mathbb{P}_{\text{S}}$ with respect to the gauge transformation induced by the edge-simplicity constraint in the discrete phase space, whose infinitely small transformation is generated by \eqref{edgesimtrans2}. Then, the action of $\mathbb{P}_{\text{S}}$ on the constraint surface defined by edge-simplicity constraint can be given as
\begin{equation}\label{qgr0}
\mathbb{P}_{\text{S}}\circ h_e :=\int_{SO(D-1)}d\bar{g}\left(u_ee^{\xi^o\tau_o}(\bar{g}e^{\bar{\xi}^\mu_e\bar{\tau}_\mu})\tilde{u}_e^{-1}\right) =h^{s}_e,
\end{equation}
\begin{equation}\label{qgr00}
\mathbb{P}_{\text{S}}\circ X_e =X_e,
\end{equation}
where we used that $h_e=u_ee^{\xi\tau_o}e^{\bar{\xi}^\mu_e\bar{\tau}_\mu}\tilde{u}_e^{-1}$, $\bar{g}\in SO(D-1)\subset SO(D+1)$, and $h^{s}_e$ is the simplicity reduced holonomy defined by
\begin{equation}
h^{s}_e=u_ee^{\xi\tau_o}\mathbb{I}^s\tilde{u}_e^{-1},
\end{equation}
where $(\mathbb{I}^s)^I_{\ J}:=(\delta_1)^I(\delta_1)_J+(\delta_2)^I(\delta_2)_J$.
Recall the simplicity reduced connection $A^S_{aIJ}:= A_{aIJ}-\beta\bar{K}_{aIJ}$ constructed in connection phase space, we can establish the following correspondence $A^S_{aIJ}$ and the simplicity reduced holonomy $h^s_e$,
  $$\xymatrix{}
\xymatrix@C=3.5cm{
(A_{aIJ},\pi^{bKL})
\ar[d]_{\textrm{(1)}} \ar[r]_{\textrm{regularization}}&(h_e,X_e) \ar[d]^{\textrm{(2)}}\\
(A^S_{aIJ},\pi^{bKL})|_{S^{abIJKL}=0}\ar[r]^{\textrm{ correspondence}}& (h^s_e,X_e)|_{S_e=0, S_v=0}}$$
where in steps (1) and (2) the symplectic reduction with respect to simplicity constraint are proceeded.
However, though the simplicity reduced holonomy $h^s_e$ and the simplicity reduced connection $A^S_{aIJ}$ has above correspondence relation, $h^s_e$ is not the holonomy defined by $A^S_{aIJ}$. This can be seen by considering the continuous limit  of $h^s_e$, which reads
\begin{equation}\label{hseconlim}
h^s_e=u_ee^{\xi\tau_o}\mathbb{I}^s\tilde{u}_e^{-1} ={u}_ee^{\xi^o_e\tau_o}\mathbb{I}^s{u}_e^{-1}h_e^{\Gamma}\simeq ({u}_e\mathbb{I}^s{u}_e^{-1}+\beta K^{\perp}_e)(\mathbb{I}+\Gamma_e),
\end{equation}
where the appearance of $\mathbb{I}^s$ leads  that $h^s_e$ is not the holonomy defined by $A^S_{aIJ}$. In fact, notice that the disappearance of $e^{\bar{\xi}^\mu_e\bar{\tau}_\mu}$ in $h^s_e$ not only reduces the gauge degrees of freedom of $\check{\xi}^\mu_e$ corresponding to $\bar{K}_{aIJ}$, but also the degrees of freedom of $\bar{\zeta}^\mu_e$ corresponding to some components of $\Gamma_{aIJ}$. Besides, by using \eqref{hseconlim}, one can also check that $h^s_\alpha$ constructed on a loop $\alpha$ can not capture the degrees of freedom of the intrinsic curvature, while it is able to capture the degrees of freedom of the extrinsic curvature properly, see more details in Appendix \ref{app1}. Thus, the variables constructed based on $h^s_\alpha$ have different interpretation as that based on $h_\alpha$. Finally, we can conclude that the regularization is not commutative to the gauge reduction with respect to simplicity constraint, this point can be regarded as another aspect of the anomaly of the simplicity constraint.

Since the construction of kinds of operators in quantum theory of all dimensional LQG relies on the regularized formulation of the simplicity reduced connection $A^S_{aIJ}$, it is worth to construct the holonomy corresponds to $A^S_{aIJ}$. Let us define
\begin{equation}\label{heS}
({h}_e^S)^I_{\ L}:=(h^s_e)^I_{\ L}+((\mathbb{I})^I_{\ J}+V_e^{IK}V_{e,KJ})(h^{\Gamma}_e)^{J}_{\ L}=u_ee^{\xi\tau_o}e^{\bar{\zeta}_e^\mu\bar{\tau}_\mu}\tilde{u}_e^{-1}
 \end{equation}
 on the gauge reduced holonomy-flux phase space with respect to edge-simplicity constraint.
One can check that
 \begin{equation}
{h}^S_e ={u}_ee^{\xi_e\tau_o}{u}_e^{-1}h_e^{\Gamma}\simeq (\mathbb{I}+\beta K^{\perp}_e)(\mathbb{I}+\Gamma_e)
\end{equation}
in continuum limit. It is easy to see that $h^S_e$ captures the physical degrees of freedom of both intrinsic and extrinsic curvature properly and it can be regarded as the holonomy of  $A^S_{aIJ}$. We conclude this point as
  $$\xymatrix{}
\xymatrix@C=3.5cm{
(A^S_{aIJ},\pi^{bKL})|_{S^{abIJKL}=0}\ar[r]^{\textrm{ regularization}}& (h^S_e,X_e)|_{S_e=S_v=0}} $$
One should notice that the definition \eqref{heS} of $h^S_e$ only holds for the elementary edges $e$ of $\gamma$ whose dual lattice gives a D-polytope partition of $\sigma$. For a loop $\alpha=e_1\circ e_2\circ ...\circ e_n$ with $e_1, e_2,..., e_n$ being the elementary edges of $\gamma$, we have $h^S_{\alpha}:=h^S_{e_1} h^S_{e_2} ...h^S_{ e_n}$.
 As we will see in section \ref{sec5}, the properties of $h^s_e$ and $h^S_e$ will be key ingredients in the construction of the scalar constraint operator.

\section{Quantum gauge reduction with respect to simplicity constraint}\label{sec3}
\subsection{The solution space of quantum simplicity constraint}\label{sec3.1}
The Hilbert space $\mathcal{H}$ of all dimensional LQG is given by the completion of the space of cylindrical functions on the quantum configuration space, which can be decomposed into the sectors --- the Hilbert spaces associated to graphs. For a given graph $\gamma$ with $|E(\gamma)|$ edges, the related Hilbert space is given by $\mathcal{H}_\gamma=L^2((SO(D+1))^{|E(\gamma)|}, d\mu_{\text{Haar}}^{|E(\gamma)|})$. This Hilbert space associates to the classical phase space $\times_{e\in\gamma}T^\ast SO(D+1)_e$ aforementioned. A basis of this space is given by the spin-network functions constructed on $\gamma$ which are labelled by (1) an $SO(D+1)$ representation $\Lambda$ assigned to each edge of $\gamma$; and (2) an intertwiner $i_v$ assigned to each vertex $v$ of $\gamma$. Then, each basis state $\Psi_{\gamma,{\vec{\Lambda}}, \vec{i}}(\vec{h}_e)$, as a wave function on $\times_{e\in\gamma}SO(D+1)_e$, can be given by
\begin{eqnarray}
\Psi_{\gamma,{\vec{\Lambda}}, \vec{i}}(\vec{h}(A))\equiv \bigotimes_{v\in\gamma}{i_v}\,\, \rhd\,\, \bigotimes_{e\in\gamma} \pi_{\Lambda_e}(h_{e}(A)),
\end{eqnarray}
where $\vec{h}(A):=(...,h_e(A),...), \vec{\Lambda}:=(...,\Lambda_e,...), e\in\gamma$, $\vec{i}:=(...,i_v,...), v\in\gamma$ , $\pi_{\Lambda_e}(h_{e})$ denotes the matrix of holonomy $h_e$ associated to edge $e$ in the representation labelled by $\Lambda_e$, and $\rhd$ denotes the contraction of  the representation matrixes of holonomies with the intertwiners. Hence, the wave function $\Psi_{\gamma,{\vec{\Lambda}}, \vec{i}}(\vec{h}(A))$ is simply the product of the functions on $SO(D+1)$, which are given by specified components of the holonomy matrices selected by the intertwiners at the vertices.
The action of the elementary operators---holonomy operator and flux operator---on the spin-network functions can be given as
\begin{eqnarray}
 \hat{ h}_{e}(A)\circ \Psi_{\gamma,{\vec{\Lambda}}, \vec{i}}(\vec{h}(A)) &=& { h}_e(A) \Psi_{\gamma,{\vec{\Lambda}}, \vec{i}}(\vec{h}(A)) \\\nonumber
  \hat{F}_e^{IJ}\circ\Psi_{\gamma,{\vec{\Lambda}}, \vec{i}}(\vec{h}(A)) &=&-\mathbf{i}\hbar\kappa\beta R_e^{IJ}\Psi_{\gamma,{\vec{\Lambda}}, \vec{i}}(\vec{h}(A))
\end{eqnarray}
where the holonomy operator acts by multiplying,  $R_{e}^{IJ}:=\text{tr}((\tau^{IJ}h_e)^{\text{T}}\frac{\partial}{\partial h_e})$  is the right
invariant vector fields on $SO(D+1)$ associated to the edge $e$, and $\text{T}$ denoting the transposition of the matrix. Then, the other operators in all dimensional LQG, such as spatial geometric operators and scalar constraint operators, can be constructed based on these elementary operators \cite{long2020operators,Long:2020agv,Zhang:2015bxa}.

Now one can proceed the quantum gauge reduction procedures to obtain the kinematic physical Hilbert space. To achieve this goal, one needs to solve the kinematic constraints, including Gaussian constraint, edge-simplicity constraint and vertex-simplicity constraint in $\mathcal{H}$. Following the results given in Sec.\ref{sec2.2}, the Gaussian constraint and edge-simplicity constraint are imposed strongly and the corresponding solution space is spanned by the edge-simple and gauge invariant spin-network states, which are constructed by assigning simple representations of $SO(D+1)$ to edges and gauge invariant intertwiners to vertices of the associated graphes. Besides, the anomalous vertex simplicity constraints are imposed weakly and the corresponding weak solutions are given by the spin-network states labelled by the simple coherent intertwiners at vertices \cite{long2019coherent}. Specifically, a typical spin-network state labelled by the gauge invariant simple coherent intertwiners at vertices is given by
\begin{equation}\label{scinter}
\Psi_{\gamma,\vec{N},\vec{\mathcal{I}}_{\text{s.c.}}}(\vec{h}(A))=\text{tr}(\otimes_{e\in\gamma} \pi_{N_e}(h_e(A))\otimes_{v\in\gamma}\mathcal{I}_v^{\text{s.c.}})
\end{equation}
where $\pi_{N_e}(h_e(A))$ denotes the representation matrix of $h_e(A)$ with $N_e$ being an non-negative integer labeling a simple representation of $SO(D+1)$, and $\vec{\mathcal{I}}_{\text{s.c.}}$ is defined by $\vec{\mathcal{I}}_{\text{s.c.}}:=(...,\mathcal{I}_v^{\text{s.c.}},...)$ with $\mathcal{I}_v^{\text{s.c.}}$ being the so-called gauge invariant simple coherent intertwiner labeling the vertex $v\in\gamma$ \cite{long2019coherent}. More explicitly, the gauge invariant simple coherent intertwiner is defined as
 \begin{equation}
 \mathcal{I}_v^{\text{s.c.}}:=\int_{SO(D+1)}dg\otimes_{e: b(e)=v}\langle N_e,V_e|g
 \end{equation}
 where all the edges linked to $v$ are re-oriented to be outgoing at $v$ without loss of generality, the labels $V_e$ satisfies the classical vertex-simplicity constraint as
 \begin{equation}
 V_e^{[IJ}V_{e'}^{KL]}=0,\quad \forall\  b(e)=b(e')=v,
 \end{equation}
 and $|N_e,V_e\rangle$ is the Perelomov type coherent state of $SO(D+1)$ in the simple representation space labelled by $N_e$ \cite{Long:2020euh}, which satisfies
 \begin{equation}
\langle N_e,V_e| \tau^{IJ}|N_e,V_e\rangle=\mathbf{i}N_eV_e^{IJ}.
 \end{equation}

It is ready to relate the procedures of classical reduction with respect to simplicity constraint to the quantum case. Notice that the quantum theory is based on the holonomy-flux variables, so that we follow the reduction procedures introduced by the twisted geometry parametrization of holonomy-flux phase space. The key step in this procedures is the weakly imposition of quantum vertex-simplicity constraint. Such a treatment relies on the  spin-network states labelled by the simple coherent intertwiners at vertices, which give the expectation value of flux operator by their labels with minimal uncertainty. Based on the fact that vertex simplicity constraint operators are purely composed by flux operators, we construct the simple coherent intertwiners labelled by the points on the constraint surface of both edge and vertex-simplicity constraint, such that the spin-network states labelled by the simple coherent intertwiners at vertices weakly solve the quantum vertex simplicity constraints with minimal uncertainty \cite{long2019coherent}. With this key step being completed, we can realize the whole quantum reduction procedures and give the correspondence between the classical and quantum reductions, which can be illustrated as follows.
\begin{equation}\label{flowchart222}
\xymatrix{}
\xymatrix@C=3.5cm{
\times_{e\in\gamma}T^\ast SO(D+1)_e
\ar[d] \ar[r]_{\textrm{quantization }}&\mathcal{H}_\gamma \ar[d]^{\textrm{(i)}}\\
\times_{e\in\gamma}\check{P}_e\ar[d]\ar[r]^{\textrm{ quantization}}& \mathcal{H}^{s}_\gamma \ar[d]^{\textrm{(ii)}}\\
\check{H}_\gamma\ar[d] \ar[r]^{\textrm{ quantization}}& \mathcal{H}^{s}_{\gamma,\text{inv.}} \ar[d]^{\textrm{(iii)}} \\
\check{H}_\gamma^{\text{s.}} \ar[r]^{\textrm{ quantum}}_{\textrm{corespondence}}& \mathcal{H}^{S}_{\gamma,\text{inv.}}  }
\end{equation}
where the procedures at left-hand side is just the flow chart \eqref{flowchart111}, and the procedures at left-hand side are explained as follows. In step (i), the edge-simplicity constraint is imposed strongly, and we get the cylindrical function space $\mathcal{H}_\gamma^s$ spanned by the spin-network functions whose edges are labelled by simple representations of $SO(D+1)$. Then in the step (ii), we impose the quantum Gauss constraint which further restricts the state space $\mathcal{H}_\gamma^s$ to the gauge (with respect to Gaussian constraint) invariant subspace $\mathcal{H}_{\gamma, \text{inv.}}^s$. In the key step (iii), we weakly impose the vertex-simplicity constraint based on the spin-network basis labelled by the coherent intertwiners, and it leads to the kinematical physical Hilbert space $\mathcal{H}^{S}_{\gamma,\text{inv.}}$ of all dimensional LQG. The resulted space $\mathcal{H}^{S}_{\gamma,\text{inv.}}$ is spanned by the spin-network states $T_{\gamma,\vec{N},\vec{\mathcal{I}}_{\text{s.c.}}}(\vec{h}_e(A))$ defined by Eq.\eqref{scinter}. Thus, the procedures in the right-hand side of \eqref{flowchart222} faithfully realizes the classical reduction procedures in the left-hand side of \eqref{flowchart222} in the quantum state space, up to some quantum perturbations of the weakly vanished vertex-simplicity constraint operators.
\subsection{Quantum gauge reduction with respect to simplicity constraint}\label{sec3.2}
In order to realize the quantum gauge reduction with respect to simplicity constraint, one need to establish the gauge (with respect to simplicity constraint) invariant holonomy and flux operators, and they will be referred to as simplicity reduced holonomy and flux operators in the following part of this paper. Since the gauge transformations with respect to simplicity constraint are generated by edge-simplicity constraint in holonomy-flux phase space, let us consider the construction of simplicity reduced holonomy and flux operators in the solution space $\mathcal{H}^{s}_{\gamma}$ of quantum edge-simplicity constraint. Then, it is easy to check that the flux operator $\hat{X}_e$ satisfies
\begin{equation}
[\hat{X}^{MN}_e, \hat{X}^{[IJ}_e\hat{X}^{KL]}_e]\circ f_{\gamma}=0, \quad \text{for} \ f_\gamma\in \mathcal{H}^{s}_{\gamma}\ \text{and}\  e\in\gamma,
\end{equation}
while the holonomy operator $\hat{h}_e$ gives
\begin{equation}
[\hat{h}_e, \hat{X}^{[IJ}_e\hat{X}^{KL]}_e]\circ f_{\gamma}\neq0, \quad \text{for} \ f_\gamma\in \mathcal{H}^{s}_{\gamma}\ \text{and}\  e\in\gamma
\end{equation}
generally, where $\hat{X}_e^{[IJ}\hat{X}_e^{KL]}$ is the edge-simplicity constraint operator which induces the gauge transformation with respect to the simplicity constraint in quantum theory. Thus, the flux operator $\hat{X}_e$ is simplicity reduced in $\mathcal{H}^{s}_{\gamma}$ while the holonomy operator $\hat{h}_e$ is not.
To find the simplicity reduced holonomy operator, let us define
 a  projection operator $\widehat{\mathbb{P}}_{\text{S}}$ which projects an arbitrary quantum state into the solution space $\mathcal{H}^{s}_{\gamma}$ of edge-simplicity constraint. Then, one can check that
\begin{equation}
[\widehat{\mathbb{P}}_{\text{S}}\hat{h}_e\widehat{\mathbb{P}}_{\text{S}}, \hat{X}^{[IJ}_e\hat{X}^{KL]}_e]\circ f_{\gamma}=0, \quad \text{for} \ f_\gamma\in \mathcal{H}^{s}_{\gamma}\ \text{and}\  e\in\gamma.
\end{equation}
Thus,  the simplicity reduced holonomy operator  $\widehat{h^{s}_e}$ can be defined as
 \begin{equation}\label{simholo}
\widehat{h^{s}_e}:=\widehat{\mathbb{P}}_{\text{S}}\hat{h}_e\widehat{\mathbb{P}}_{\text{S}}.
\end{equation}
One should note that the action of $\widehat{\mathbb{P}}_{\text{S}}$ on quantum state is distinguished with the action of ${\mathbb{P}}_{\text{S}}$ on the classical variables $(h_e, X_e)$. To understand this point, notice that the infinitely small transformation of $\widehat{\mathbb{P}}_{\text{S}}$ is generated by the action of edge-simplicity constraint operator on a cylindrical function $f_\gamma(...,h_e,...)$, which reads
\begin{equation}\label{poissonRRA}
\hat{X}_e^{[IJ}\hat{X}_e^{KL]}\circ f_\gamma(...,h_e,...)\propto \{X_e^{[IJ},\{X_e^{KL]}, f_\gamma(...,h_e,...)\}\}\propto f_\gamma (...,\tau^{[IJ}\tau^{KL]}h_e,...),
\end{equation}
wherein the transformation of holonomy is not identical with the transformation of holonomy induced by the classical edge-simplicity constraint given in Eq.\eqref{esimtrans}. Finally, we conclude the gauge reduction with respect to the simplicity constraint in both classical and quantum theory with the following corresponding table:
\\

\begin{tabular}{|l|c|c|c|}
\hline
 & edge-simplicity constraint& gauge invariant state & simplicity reduced variables\\
 \hline
 classical& ${X}_e^{[IJ}{X}_e^{KL]} \approx 0$ &$({h^{s}_e}, X_e)|_{{X}_e^{[IJ}{X}_e^{KL]} =0}$&$({h^{s}_e}, X_e)|_{{X}_e^{[IJ}{X}_e^{KL]} =0}$\\
 \hline
quantum & $\hat{X}_e^{[IJ}\hat{X}_e^{KL]}\circ f_\gamma=0$ & $f_\gamma\in \mathcal{H}_\gamma^s$ & $(\widehat{h^{s}_e},\hat{X}_e)$ \text{with\ domain} $\mathcal{H}_\gamma^s$ \\
\hline
\end{tabular}
\\ \\
Note that the quantum simplicity reduced variables $(\widehat{h^{s}_e},\hat{X}_e)$ \text{with\ domain} $\mathcal{H}_\gamma^s$ are  constructed by projecting the action of holonomy and flux operators $(\widehat{h_e},\hat{X}_e)$ into $\mathcal{H}_\gamma^s$, instead of quantizing the classical  simplicity reduced variables $({h^{s}_e}, X_e)|_{{X}_e^{[IJ}{X}_e^{KL]} =0}$ directly. Thus, one may doubt whether $(\widehat{h^{s}_e},\hat{X}_e)$ \text{with\ domain} $\mathcal{H}_\gamma^s$ can be regarded as the quantization of $({h^{s}_e}, X_e)|_{{X}_e^{[IJ}{X}_e^{KL]} =0}$. In next subsection, we will give the reasonableness of this argument by showing that $(\widehat{h^{s}_e},\hat{X}_e)$ \text{with\ domain} $\mathcal{H}_\gamma^s$ reproduce $({h^{s}_e}, X_e)|_{{X}_e^{[IJ}{X}_e^{KL]} =0}$ in the semi-classical limit.

However, as aforementioned in Sec.\ref{2.3}, since the simplicity reduced holonomy $h^s_e$ is not able to capture the degrees of freedom of the intrinsic and extrinsic curvature properly, its quantum operator $\widehat{h^s_e}$ can not be used to construct the operators involved curvatures directly. Thus, it is also worth to construct the operator $\widehat{h^S_e}$ corresponding to the holonomy $h^S_e$ of $A^S_{aIJ}$. Recall
  \begin{equation}
({h}_e^S)^I_{\ L}:=(h^s_e)^I_{\ L}+((\mathbb{I})^I_{\ J}+V_e^{IK}V_{e,KJ})(h^{\Gamma}_e)^{J}_{\ L}
 \end{equation}
 defined on the reduce holonomy-flux phase space with respect to edge-simplicity constraint.
It is easy to see that we still need to construct the operators corresponding to $V_e$ and $h^{\Gamma}_e$.  Notice that $V^{IJ}=\frac{2X^{IJ}}{\sqrt{2X_{KL}X^{KL}}}$ holds on the edge-simplicity constraint surface, thus we have
 \begin{equation}
V_e^{IK}V_{e,KJ}=\frac{2X_e^{IK}X_{e,KJ}}{X_{e,MN}X_e^{MN}}
\end{equation}
and it can be quantized as a function of flux operator acting in the space $\mathcal{H}_\gamma^s$, which reads
 \begin{equation}
\hat{V}_e^{IK}\hat{V}_{e,KJ}=2\hat{X}_e^{IK}\hat{X}_{e,KJ}(\hat{X}_{e,MN}\hat{X}_e^{MN})^{-1},
\end{equation}
where $(\hat{X}_{e,MN}\hat{X}_e^{MN})^{-1}$ is the inverse operator of $\hat{X}_{e,MN}\hat{X}_e^{MN}$. It is easy to see that $\hat{X}_{e,MN}\hat{X}_e^{MN}$ acts as the Casimir operator of $SO(D+1)$ and it has discrete eigen-spectrum. Thus, the inverse operator of $\hat{X}_{e,MN}\hat{X}_e^{MN}$ can be defined as
\begin{equation}\label{EKKK}
(\hat{X}_{e,MN}\hat{X}_e^{MN})^{-1}:=\sum E^{-1}_k|k\rangle\langle k|,
\end{equation}
where $|k\rangle$ represents the eigen state of $\hat{X}_{e,MN}\hat{X}_e^{MN}$ with $E_k$ being the corresponding eigenvalue, and the summation takes over all of $|k\rangle$ with $E_k\neq0$.
 Then, the major obstacle to construct the operator  corresponding to $h^{S}_e$ is the quantization of $h^{\Gamma}_e$. Note that $h^{\Gamma}_e$ is the holonomy of the spin connection $\Gamma_{aIJ}$ determined by $\pi^{aIJ}$. Thus, it is reasonable to define the smeared spin connection operator $\widehat{\Gamma}_e:=\Gamma_e(\hat{X})$ as a function of $\hat{X}_e$, and then the operator  corresponding to $h^{\Gamma}_e$ can be given by $\widehat{h^{\Gamma}_e}:=\exp(\widehat{\Gamma}_e)$. However, $\Gamma_{aIJ}$ is a rather complicated function of $\pi^{bKL}$ so that the construction of  $\widehat{\Gamma}_e=\Gamma_e(\hat{X})$ is a knotty issue (see a related research in Ref.\cite{Lewandowski:2021iun}), and we will leave it to future study.

\subsection{Realization of quantum gauge reduction based on twisted geometry coherent state}\label{sec3.3}
To show the semi-classical property of the simplicity reduced operators with respect to the twisted geometry coherent state,  it is sufficient to consider the phase space and the Hilbert space associated to a single edge $e$. Then, we have the simplicity reduced operators $(\widehat{h^{s}_e},\hat{X}_e)$ \text{with\ domain} $\mathcal{H}_e^s$ and the twisted geometry coherent state $\breve{\Psi}_{\mathbb{H}^o_e}\in \mathcal{H}_e^s$ labelled by the twisted geometry parameters $\mathbb{H}^o_e:=(\eta_e,\xi_e,V_e,\tilde{V}_e)$ on the edge $e$, where the semi-classicality parameter $t$ is defined by $t:=\frac{\kappa\hbar}{a^{D-1}}$. Denoted by $\phi^t_{\mathbb{H}^o_e}:=\frac{\breve{\Psi}_{\mathbb{H}^o_e}}{||\breve{\Psi}_{\mathbb{H}^o_e}||}$ the normalized twisted geometry coherent state, then the semi-classical property of $\widehat{h^{s}_e}$ and $\hat{X}_e$ can be shown by evaluating their expectation values and matrix elements  in the twisted geometry coherent state basis. This calculations have been done in \cite{Long:2021lmd}\cite{Long:2022cex}\cite{Long:2021xjm} and it shows that the expectation values and matrix elements of $\widehat{h^{s}_e}$ and $\hat{X}_e$ are well-estimated by their classical correspondence $({h^{s}_e}, X_e)|_{{X}_e^{[IJ}{X}_e^{KL]} =0}$. More explicitly, notice that
\begin{equation}
\langle\phi^t_{\mathbb{H}^o_e}|\widehat{h^{s}_e}|\phi^t_{\mathbb{H}^o_e}\rangle =\langle\phi^t_{\mathbb{H}^o_e}|\widehat{\mathbb{P}}_{\text{S}}\hat{h}_e\widehat{\mathbb{P}}_{\text{S}} |\phi^t_{\mathbb{H}^o_e}\rangle=\langle\phi^t_{\mathbb{H}^o_e}|\hat{h}_e |\phi^t_{\mathbb{H}^o_e}\rangle,
\end{equation}
then the expectation values of $\widehat{h^{s}_e}$ and $\hat{X}_e$ are evaluated by
 \begin{eqnarray}\label{exFfinal}
\langle\phi^t_{\mathbb{H}^o_e}|\hat{X}_e^{IJ} |\phi^t_{\mathbb{H}^o_e}\rangle \stackrel{\text{large}\ \eta_e}{=}\frac{\eta_e}{2}V^{IJ}(1+\mathcal{O}(t)),
\end{eqnarray}
      \begin{eqnarray}\label{exHfinal}
\langle\phi^t_{\mathbb{H}^o_e}|\widehat{u_e^{-1}h_e\tilde{u}_e}  |\phi^t_{\mathbb{H}^o_e}\rangle \stackrel{\text{large}\ \eta_e}{=}u_e^{-1}h^s_e\tilde{u}_e(1+\mathcal{O}(t))
\end{eqnarray}
and the matrix elements of $\widehat{h^{s}_e}$ and $\hat{X}_e$ are evaluated by
      \begin{eqnarray}\label{Ffinal}
\left|\langle\phi^t_{\mathbb{H}^o_e}|\hat{X}_e^{IJ} |\phi^t_{\mathbb{H}'^o_e}\rangle -\frac{\eta'_e}{2}V'^{IJ}\langle\phi^t_{\mathbb{H}^o_e} |\phi^t_{\mathbb{H}'^o_e}\rangle\right|
&\stackrel{\text{large}\ \eta_e}{\lesssim}&t\left|f_{X}(\mathbb{H}^o_e,\mathbb{H}'^o_e)\right|\cdot\left|\langle\phi^t_{\mathbb{H}^o_e} |\phi^t_{\mathbb{H}'^o_e}\rangle\right|,
\end{eqnarray}
      \begin{eqnarray}\label{Hfinal}
\left|\langle\phi^t_{\mathbb{H}^o_e}|\widehat{u'^{-1}_eh_e\tilde{u}'_e}  |\phi^t_{\mathbb{H}'^o_e}\rangle -u'^{-1}_eh'^s_e\tilde{u}'_e\langle\phi^t_{\mathbb{H}^o_e} |\phi^t_{\mathbb{H}'^o_e}\rangle\right|
&\stackrel{\text{large}\ \eta_e}{\lesssim}&t\left|f_{h}(\mathbb{H}^o_e,\mathbb{H}'^o_e)\right|\cdot\left|\langle\phi^t_{\mathbb{H}^o_e} |\phi^t_{\mathbb{H}'^o_e}\rangle\right|,
\end{eqnarray}
where $f_{X}(\mathbb{H}^o_e,\mathbb{H}'^o_e)$ and $f_{h}(\mathbb{H}^o_e,\mathbb{H}'^o_e)$  are functions which are always suppressed by the exponentially decayed factor $\left|\langle\phi^t_{\mathbb{H}^o_e} |\phi^t_{\mathbb{H}'^o_e}\rangle\right|$ for $\mathbb{H}^o_e\neq\mathbb{H}'^o_e$, and $u'_e, \tilde{u}'_e$ in the holonomy operator $\widehat{u'^{-1}_eh_e\tilde{u}'_e}$ act on the basis vectors which select specific matrix element of the holonomy in the definition representation of $SO(D+1)$.

\section{On the construction of quantum scalar constraint in all dimensional LQG}\label{sec5}
The simplicity reduced holonomy $h^s_e$ takes a different geometric interpretation from the original holonomy $h_e$. Hence, the operators whose constructions involve holonomies should be considered carefully, to ensure they take the correct geometric interpretations. In this section, we will consider the construction of the scalar constraint operator. As we will see, since the appearance of the simplicity reduced holonomy $h^s_e$, the standard strategy  fails to give a correct scalar constraint operator in all dimensional LQG. To overcome this problem, we will propose three new strategies for constructing the scalar constraint operator, which point out the direction of further researches on the dynamics of all dimensional LQG. Notice that our discussions only focus on the factors involving holonomies in the scalar constraint, thus the studies of the factors composed by fluxes are omitted and one can find the details in the Refs.\cite{Bodendorfer:Qu}\cite{Zhang:2015bxa}.
\subsection{The problematic standard strategy}\label{sec5.1}
The quantum gauge reduction with respect to the simplicity constraint introduced in above sections helps us to construct  the scalar constraint operator in all dimensional LQG.
Similar to the $SU(2)$ connection  formulation of (1+3)-GR, one may establish the scalar constraint in $SO(D+1)$ connection  formulation of (1+D)-dimensional GR with two terms---the so called Euclidean term $C_{\text{E}}$ and Lorentzian term $C_{\text{L}}$ \cite{Bodendorfer:Qu}. The Euclidean term $C_{\text{E}}$ reads
\begin{equation}
C_{\text{E}}:=\frac{1}{\sqrt{\det(q)}}F_{abIJ}\pi^{aIK}{\pi^{b\ J}_K}
\end{equation}
with $
F_{abIJ}:=\partial_aA_{bIJ}-\partial_bA_{aIJ}+\eta^{KL}A_{aIK}A_{bLJ}-\eta^{KL}A_{aJK}A_{bIJ}
$. Define
\begin{equation}
 C_{\text{E}}[1]:=\int_{\sigma}d^Dy C_{\text{E}}(y),
 \end{equation}
then the Lorentzian term $C_{\text{L}}$ reads
\begin{eqnarray}
C_{\text{L}}&:=&
-\frac{8(1+\beta^2)}{\sqrt{\det(q)}}
K_{[a|I|}K_{b]J}E^{aI}E^{b J}\\\nonumber
&=&\frac{4(1+\beta^2)}{\sqrt{\det(q)}}
[K_{b}^{\  a}K_{a}^{\ b}-K^2],
\end{eqnarray}
where $K(x):=K_{aI}(x)E^{aI}(x)$ and $K_{b}^{\  a}:=K_{bI}E^{aI}$ satisfy
\begin{equation}
K(x)=-\frac{1}{4\kappa\beta^2}\{C_{\text{E}}(x),V(x,\epsilon)\}
\end{equation}
and
\begin{equation}
K_{aI}(x)E^{bI}(x)=-\frac{1}{8\kappa^2\beta^3}\pi^{bKL}(x)\{A_{aKL}(x), \{C_{\text{E}}[1],V(x,\epsilon)\}\}
\end{equation}
 on the constraint surface of both Gaussian and simplicity constraint,
 with $R(x,\epsilon)\ni x$ being a D-dimensional hyper-cube with coordinate scale $\epsilon$ and $V(x,\epsilon)$ being the volume of $R(x,\epsilon)$.
One can check that  $H_{\text{E}}$ contains the pure gauge component $\bar{K}_{aIJ}$ through the following identity
\begin{eqnarray}
C_{\text{E}}&:=&\frac{1}{\sqrt{\det(q)}}F_{abIJ}\pi^{aIK}{\pi^{b\ J}_K}\\\nonumber
&=&-\sqrt{\det(q)}R-\frac{\beta^2}{\sqrt{\det(q)}}(4[K_{b}^{\ a}K_a^{\ b}-K^2]+(\bar{K}_{bIK}E^{aI})(\bar{K}_{aJ}^{\ \ \, K}E^{bJ})),
\end{eqnarray}
which holds on the constraint surface of both Gaussian and simplicity constraint, where $R$ is the scalar curvature of $\Gamma_{aIJ}$ defined by
 \begin{equation}\label{scalarR}
   R:=-\frac{1}{\det(q)}R_{abIJ}\pi^{aIK}{\pi^{b\ J}_K}
 \end{equation}
 with $R_{abIJ}:=\partial_a\Gamma_{bIJ}-\partial_b\Gamma_{aIJ}+\eta^{KL}\Gamma_{aIK}\Gamma_{bLJ} -\eta^{KL}\Gamma_{aJK}\Gamma_{bIJ}
$. Thus, in order to get the correct gauge invariant ADM scalar constraint on the constraint surface of both Gaussian and simplicity constraint,  the scalar constraint in $SO(D+1)$ connection  formulation of (1+D)-GR must contain an additional term $\frac{\beta^2}{\sqrt{\det(q)}}(\bar{K}_{bIK}E^{aI})(\bar{K}_{aJ}^{\ \ \, K}E^{bJ})$  to cancel the gauge variant term in $C_{\text{E}}$, the final scalar constraint  reads
\begin{equation}\label{scalar1}
C=C_{\text{E}}+C_{\text{L}}+\frac{\beta^2}{\sqrt{\det(q)}}(\bar{K}_{bIK}E^{aI})(\bar{K}_{aJ}^{\ \ \, K}E^{bJ}).
\end{equation}

Comparing with the $SU(2)$ loop quantum gravity in (1+3)-dimension, the additional term $\frac{\beta^2}{\sqrt{\det(q)}}(\bar{K}_{bIK}E^{aI})(\bar{K}_{aJ}^{\ \ \, K}E^{bJ})$ introduce a huge obstacle to regularize and quantize the scalar constraint in all dimensional LQG. By projecting the covariant derivation of $\pi^{aIJ}$ properly, the term $\frac{\beta^2}{\sqrt{\det(q)}}(\bar{K}_{bIK}E^{aI})(\bar{K}_{aJ}^{\ \ \, K}E^{bJ})$  is re-formulated as a $DF^{-1}D$ term composed by the connection variables. In fact, the $DF^{-1}D$ term is a rather complicated function of $A_{aIJ}$ and $\pi^{bIJ}$ so that its regularization and quantization are full of ambiguities.  However, the key issue is not the treatment of the $DF^{-1}D$ term when we consider the quantization of the scalar constraint. As we will see, the operator $\hat{C}_{\text{E}}$ corresponding to Euclidean term  lose its original geometric interpretation in the space $\bigoplus_{\gamma}\mathcal{H}^{s}_{\gamma}$, since the simplicity reduced holonomy which will appear in $\hat{C}_{\text{E}}$ can not give the curvatures correctly. Let us now explain this point explicitly.

Following the regularization and quantization procedures in \cite{Bodendorfer:Qu}, the Euclidean term $C_{\text{E}}$ and Lorentzian term $C_{\text{L}}$ can be quantized directly, which leads to
\begin{equation}\label{CECL}
\hat{C}_{\text{E}}[N]=\lim_{\epsilon\to 0}\sum_{\square\in \mathfrak{P}}\hat{C}^\square_{\text{E}}[N],\quad \hat{C}_{\text{L}}[N]=\lim_{\epsilon\to 0}\sum_{\square\in \mathfrak{P}}\hat{C}^\square_{\text{L}}[N]
\end{equation}
with
\begin{equation}\label{CE}
\hat{C}^\square_{\text{E}}[N]:=N(v_\square)\cdot {}^\epsilon\!\left(\widehat{\frac{\pi^{[a|IK|}{\pi^{b]\ J}_{\ K}}}{\sqrt{\det(q)}}}\right)_{v_\square}\cdot(\hat{h}_{\alpha_{s_a,s_b}})_{[IJ]}
\end{equation}
and
\begin{eqnarray}\label{CL}
\hat{C}^\square_{\text{L}}[N]&:=&\frac{2(1+\beta^2)}{(8\kappa^2\hbar^2\beta^3)^2}N(v_\square)\cdot {}^\epsilon\!\left(\widehat{\frac{\pi^{[a|IK|}}{\sqrt[4]{\det(q)}}}\right)_{v_\square} \cdot\widehat{ (h_{s_a})}_{I}^{\ M}\left[\widehat{(h_{s_a}^{-1})}_{MK}, [\hat{C}_{\text{E}}[1],\hat{V}(v_\square,\epsilon)]\right]
\\\nonumber
&&\cdot{}^\epsilon\!\left(\widehat{\frac{{\pi^{b]JL}}}{\sqrt[4]{\det(q)}}}\right)_{v_\square}\cdot\widehat{(h_{s_b})}_{J}^{\ N}\left[\widehat{(h_{s_b}^{-1})}_{NK}, [\hat{C}_{\text{E}}[1],\hat{V}(v_\square,\epsilon)]\right],
\end{eqnarray}
where $N(x)$ is the lapse function, $\square$ denotes an elementary cell of the hyper-cubic partition $\mathfrak{P}$ of $\sigma$, $\epsilon$ represents the scale of $\square$, $v_\square$ is a vertex of $\square$, $s_a$ represents the edges of $\square$ based at $v_\square$, $\alpha_{s_a,s_b}$ represents the oriented loop based at $v_\square$ and $s_a, s_b$. Besides, the operator ${}^\epsilon\!\left(\widehat{\frac{\pi^{[a|IK|}{\pi^{b]\ J}_{\ K}}}{\sqrt{\det(q)}}}\right)_{v_\square}$ and ${}^\epsilon\!\left(\widehat{\frac{\pi^{aIK}}{\sqrt[4]{\det(q)}}}\right)_{v_\square} $ are constructed by regularizing and quantizing the factors $\frac{\pi^{aIK}{\pi^{b\ J}_K}}{\sqrt{\det(q)}}$ and $\frac{\pi^{aIK}}{\sqrt[4]{\det(q)}}$ respectively, with the regularization being compatible with the partition $\mathfrak{P}$ at $v_\square$, see more details in Ref.\cite{Bodendorfer:Qu}. Notice that the operator ${}^\epsilon\!\left(\widehat{\frac{\pi^{[a|IK|}{\pi^{b]\ J}_{\ K}}}{\sqrt{\det(q)}}}\right)_{v_\square}$ is a polynomial of $(\hat{V}(v_\square,\epsilon))^{1+x}$ and $\hat{h}_{s_a}(\hat{V}(v_\square,\epsilon))^{1+x}\hat{h}^{-1}_{s_a}$ with $x>-1$, thus it is commutative with $\widehat{\mathbb{P}}_{S}$. Then, consider a state $|\phi\rangle \in \bigoplus_{\gamma}\mathcal{H}^{s}_{\gamma}$ which satisfies
\begin{equation}
\widehat{\mathbb{P}}_{S}|\phi\rangle=|\phi\rangle,
\end{equation}
we have
\begin{equation}
\langle\phi|\hat{C}_{\text{E}}[N]|\phi'\rangle=\langle\phi|\widehat{\mathbb{P}}_{S} \hat{C}_{\text{E}}[N]\widehat{\mathbb{P}}_{S}|\phi' \rangle=\langle\phi|\hat{C}^s_{\text{E}}[N]|\phi'\rangle,
\end{equation}
where we defined
 \begin{equation}\label{CSE}
\hat{C}^s_{\text{E}}[N]=\lim_{\epsilon\to 0}\sum_{\square\in \mathfrak{P}}\hat{C}^{s,\square}_{\text{E}}[N]
\end{equation}
with
\begin{equation}
\hat{C}^{s,\square}_{\text{E}}[N]:=N(v_\square)\cdot {}^\epsilon\!\left(\widehat{\frac{\pi^{[a|IK|}{\pi^{b]\ J}_{\ K}}}{\sqrt{\det(q)}}}\right)_{v_\square}\cdot(\widehat{h^s}_{\alpha_{s_a,s_b}})_{[IJ]},
\end{equation}
which is given by replacing the holonomy operator $\hat{h}_{\alpha_{s_a,s_b}}$ in $\hat{C}_{\text{E}}[N]$ by the simplicity reduced one $\widehat{h^s}_{\alpha_{s_a,s_b}}$. By this we can conclude that $\hat{C}_{\text{E}}[N]$ is equivalent to $\hat{C}^{s}_{\text{E}}[N]$ in the space $\bigoplus_{\gamma}\mathcal{H}^{s}_{\gamma}$. Here the key point is that, when one consider the matrix element of $\hat{C}_{\text{E}}[N]$ in the space $\bigoplus_{\gamma}\mathcal{H}^{s}_{\gamma}$, the holonomy operator $\widehat{h}_{\alpha_{s_a,s_b}}$ reduces as the simplicity holonomy operator $\widehat{h^s}_{\alpha_{s_a,s_b}}$ and $\hat{C}_{\text{E}}[N]$ reduces as $\hat{C}^{s}_{\text{E}}[N]$. Note that  $\widehat{h^s}_{\alpha_{s_a,s_b}}$ corresponds to the  classical simplicity reduced holonomy $h^s_{\alpha_{s_a,s_b}}$, whose geometric interpretation is different with $h_{\alpha_{s_a,s_b}}$ . Thus, we know that the action of $\hat{C}_{\text{E}}[N]$ in $\bigoplus_{\gamma}\mathcal{H}^{s}_{\gamma}$ can not reveal the physical meaning of the classical scalar constraint $C_{\text{E}}$ at quantum level.
Besides, the Eq. \eqref{CL} is also not the operator corresponding to $C_{\text{L}}$, since its definition relies on the operator $\hat{C}_{\text{E}}[1]$.

\subsection{New strategies}\label{sec5.2}
The physical consideration involving scalar constraint must happen in the case that simplicity constraints are solved. Thus, in order to simplify the discussions, we firstly claim that the scalar constraint operators constructed in following subsections are defined in the space $\mathcal{H}^s:=\bigoplus_{\gamma}\mathcal{H}^{s}_{\gamma}$ which vanishes the edge-simplicity constraint operator.
\subsubsection{The first strategy}
Since the previous construction of the scalar constraint operator fails, one can consider some new strategies of the construction of scalar constraint operator.  In the first strategy, let us recall the simplicity reduced connection
\begin{equation}
A^S_{aIJ}\equiv A_{aIJ}-\beta\bar{K}_{aIJ}
\end{equation}
and its curvature is defined by
\begin{equation}
F^S_{abIJ}:=\partial_aA^S_{bIJ}-\partial_bA^S_{aIJ}+\eta^{KL}A^S_{aIK}A^S_{bLJ}-\eta^{KL}A^S_{aJK}A^s_{bIJ}.
\end{equation}
It is easy to check
\begin{equation}
C^S_{\text{E}}:=\frac{1}{\sqrt{\det(q)}}F^S_{abIJ}\pi^{aIK}{\pi^{b\ J}_K}=-\sqrt{\det(q)}R-\frac{4\beta^2}{\sqrt{\det(q)}}[K_{ab}K^{ab}-K^2]
\end{equation}
and
\begin{equation}
K_{aI}(x)E^{bI}(x)=-\frac{1}{8\kappa^2\beta^3}\pi^{bKL}(x)\{A_{aKL}(x), \{C^S_{\text{E}}[1],V(x,\epsilon)\}\},
\end{equation}
on the constraint surface of both Gaussian and simplicity constraint. Then, the scalar constraint can be expressed as
\begin{equation}\label{scalar2}
C=C^S_{\text{E}}+C_{\text{L}}.
\end{equation}
Now, let us consider the regularization and quantization of  $C^S_{\text{E}}$. Notice that $C^S_{\text{E}}$ takes same formulation as $C_{\text{E}}$ except that the connection $A_{aIJ}$ in $C_{\text{E}}$ is replaced by $A^S_{aIJ}$ in $C^S_{\text{E}}$. Moreover, recall that the smearing version of $A^S_{aIJ}$ is given by $h^S_e$.  Thus, following a similar  regularization and quantization procedure as that of $C_{\text{E}}$, we can give the operator corresponding to $C^S_{\text{E}}$ as
  \begin{equation}\label{CSE2}
\hat{C}^S_{\text{E}}[N]=\lim_{\epsilon\to 0}\sum_{\square\in \mathfrak{P}}\hat{C}^{S,\square}_{\text{E}}[N]
\end{equation}
with
\begin{equation}
\hat{C}^{S,\square}_{\text{E}}[N]:=N(v_\square)\cdot {}^\epsilon\!\left(\widehat{\frac{\pi^{[a|IK|}{\pi^{b]\ J}_{\ K}}}{\sqrt{\det(q)}}}\right)_{v_\square}\cdot(\widehat{h^S}_{\alpha_{s_a,s_b}})_{[IJ]},
\end{equation}
which is given by replacing the holonomy operator $\hat{h}_{\alpha_{s_a,s_b}}$ in $\hat{C}_{\text{E}}[N]$ by another holonomy operator $\widehat{h^S}_{\alpha_{s_a,s_b}}$ corresponding to the holonomy $h^S_e$ of $A^S_{aIJ}$. Accordingly, the operator corresponding to $C_{\text{L}}$ is given by
 \begin{equation}\label{CL33}
 \hat{C}_{\text{L}}[N]=\lim_{\epsilon\to 0}\sum_{\square\in \mathfrak{P}}\hat{C}^\square_{\text{L}}[N]
\end{equation}
with
\begin{eqnarray}\label{CL333}
\hat{C}^\square_{\text{L}}[N]&:=&\frac{2(1+\beta^2)}{(8\kappa^2\hbar^2\beta^3)^2}N(v_\square)\cdot {}^\epsilon\!\left(\widehat{\frac{\pi^{[a|IK|}}{\sqrt[4]{\det(q)}}}\right)_{v_\square} \cdot\widehat{ (h_{s_a})}_{I}^{\ M}\left[\widehat{(h_{s_a}^{-1})}_{MK}, [\hat{C}^S_{\text{E}}[1],\hat{V}(v_\square,\epsilon)]\right]
\\\nonumber
&&\cdot{}^\epsilon\!\left(\widehat{\frac{{\pi^{b]JL}}}{\sqrt[4]{\det(q)}}}\right)_{v_\square}\cdot\widehat{(h_{s_b})}_{J}^{\ N}\left[\widehat{(h_{s_b}^{-1})}_{NK}, [\hat{C}^S_{\text{E}}[1],\hat{V}(v_\square,\epsilon)]\right],
\end{eqnarray}
 Finally, one can conclude that the scalar constraint operator in all dimensional LQG can be given as
  \begin{equation}\label{scalarfina}
\hat{C}[N]=\hat{C}^S_{\text{E}}[N]+\hat{C}_{\text{L}}[N],
\end{equation}
where $\hat{C}^S_{\text{E}}[N]$ and $\hat{C}_{\text{L}}[N]$ are defined in Eqs.\eqref{CSE2} and \eqref{CL33} respectively.
\subsubsection{The second strategy}
In this strategy, we still consider the expression \eqref{scalar2} of scalar constraint and keep the regularization and quantization scheme for  ${C}^S_{\text{E}}$ in \eqref{scalar2}. Then, let us consider a new scheme of the regularization and quantization of ${C}_{\text{L}}$ in \eqref{scalar2}. By using Eq.\eqref{hsKK} in Appendix \ref{app1}, one can regularize ${C}_{\text{L}}$ by define
\begin{eqnarray}\label{KKsec}
&&{C}^\square_{\text{L,alt}}[N]:=\frac{8(1+\beta^2)}{\beta^2}N(v_\square)\cdot {}^\epsilon\!\left({\frac{\pi_{}^{[a|IK|}{\pi^{b]\ J}_{\ K}}}{\sqrt{\det(q)}}}\right)_{v_\square}\cdot(h^s_{\alpha_{s_a,s_b}})_{[IJ]},
\end{eqnarray}
and one can verify that
\begin{equation}
{C}_{\text{L}}[N]=\lim_{\epsilon\to0}\sum_{\square\in \mathfrak{P}}{C}^\square_{\text{L,alt}}[N]
\end{equation}
holds on the constraint surface defined by simplicity constraint. Then, the operator corresponding to  ${C}_{\text{L}}[N]$ can be given by
 \begin{equation}\label{CL44}
 \hat{C}_{\text{L}}[N]=\lim_{\epsilon\to 0}\sum_{\square\in \mathfrak{P}}\hat{C}^\square_{\text{L,alt}}[N]
\end{equation}
with
\begin{eqnarray}\label{CL444}
\hat{C}^\square_{\text{L,alt}}[N]:=\frac{8(1+\beta^2)}{\beta^2}N(v_\square)\cdot {}^\epsilon\!\left(\widehat{\frac{\pi_{}^{[a|IK|}{\pi^{b]\ J}_{\ K}}}{\sqrt{\det(q)}}}\right)_{v_\square}\cdot(\widehat{h}_{\alpha_{s_a,s_b}})_{[IJ]},
\end{eqnarray}
where the $\widehat{h^s}_{\alpha_{s_a,s_b}}$ is substituted by $\widehat{h}_{\alpha_{s_a,s_b}}$ since they are identical in the solution space of edge simplicity constraint. Finally, in this strategy, the scalar constraint operator in all dimensional LQG is given by Eq. \eqref{scalarfina} with $\hat{C}^S_{\text{E}}[N]$ and $\hat{C}_{\text{L}}[N]$ are defined in Eqs.\eqref{CSE2} and \eqref{CL44} respectively.
\subsubsection{The third strategy}
Notice that the operator $\hat{C}^S_{\text{E}}[N]$ which is involved in the first and second strategies depends on the operator $\widehat{h^S_e}$ corresponding to the holonomy $h^S_e$ of $A^S_{aIJ}$. However,  the explicit expression of $\widehat{h^S_e}$ involves another operator $\widehat{h_e^\Gamma}$ whose construction is still a difficult issue. In the third strategy,  we consider a new expression of scalar constraint to avoid the difficulty of the operator $\hat{C}^S_{\text{E}}[N]$. We can re-express the scalar constraint as
\begin{equation}\label{Cthird}
C=\frac{1}{(1+\beta^2)}C_{\text{L}}-\sqrt{\det(q)}R.
\end{equation}
By regularizing and quantizing Eq.\eqref{Cthird}, we could get a new scalar constraint operator
   \begin{equation}\label{Cthirdop}
\hat{C}[N]=\frac{1}{(1+\beta^2)}\hat{C}_{\text{L}}[N]-\hat{\tilde{R}}[N],
\end{equation}
where $\hat{C}_{\text{L}}[N]$ is given by Eq.\eqref{CL44}, and
 $\hat{\tilde{R}}[N]$ is the operator corresponding to
  \begin{equation}
  \tilde{R}[N]:=\int_{\sigma}d^DyN(y)\sqrt{\det(q)}R(y).
  \end{equation}
Notice that the operator $\hat{\tilde{R}}[N]$ has not been constructed yet in all dimensional LQG. Nevertheless, its analogue in $SU(2)$ LQG has been constructed and studied in various methods \cite{Lewandowski:2021iun}\cite{Alesci:2014aza}\cite{Assanioussi:2015gka}. It is expected to extend these methods to all dimensional LQG to give the explicit expression of $\hat{\tilde{R}}[N]$. We leave this task to further researches.

\section{Conclusion}\label{sec6}
The gauge reduction with respect to the simplicity constraint has been discussed in both classical and quantum theory of loop quantum gravity. In classical connection phase space, the symplectic reduction with respect to simplicity and Gaussian constraint can be proceeded without anomaly, which leads to the ADM phase space correctly. Different with the continuum connection theory, the simplicity constraints in discrete holonomy-flux phase space become anomalous. It has been shown that, in order to gives the discrete twisted geometry correctly, one should proceed gauge reduction  with respect to edge simplicity constraint and then impose the vertex simplicity constraint weakly, i.e. solving the vertex simplicity constraint equations. However, once we consider the gauge reduction with respect to edge simplicity constraint in holonomy-flux phase, we find that the simplicity reduced holonomy $h^s_e$ can not capture the degrees of freedom of intrinsic curvature, since its continuum limit does not reproduce the simplicity reduced connection $A^S_{aIJ}$. Besides, the holonomy operator $\hat{h}_e$ is identical with the simplicity reduced holonomy operator $\widehat{h^s_e}:=\widehat{\mathbb{P}}_{\text{S}}\hat{h}_e\widehat{\mathbb{P}}_{\text{S}}$ in the space $\mathcal{H}^s$ spanned by the states vanishing edge simplicity constraint, which means the classical correspondence of $\hat{h}_e$ acting in $\mathcal{H}^s$ is given by $h^s_e$ instead of $h_e$. This result leads that the standard strategy fails to give a correct scalar constraint operator in all dimensional LQG. Hence, we propose three new strategies to construct the scalar constraint operator in Sec.\ref{sec5.2}. In the first strategy, we establish the holonomy $h^S_e$ of the simplicity reduced connection $A^S_{aIJ}$ which captures degrees of freedom of the intrinsic and extrinsic curvature correctly, and then the operator $\widehat{h^S_e}$ is used to substitute $\hat{h}_e$ to construct the scalar constraint.  In the second strategy, we consider an alternative of the Lorentzian part of the scalar constraint based on the simplicity reduced holonomy operator $\widehat{h^s_e}$. In the third strategy, a new scheme for treating the scalar curvature term in scalar constraint is considered.
Generally, the issues introduced by the gauge reduction with respect to simplicity constraint are discussed in this paper, and several strategies are proposed to deal with them. Nevertheless, these strategies still need further studies. As we have mentioned before, $\widehat{h^S_e}$ involved in first and second strategies contains the operator which corresponds to the holonomy of Levi-Civita connection, and this operator  has not been constructed yet in all dimensional LQG. Besides, though the scalar curvature operator involved in third strategy has been established in $SU(2)$ LQG, we still need to generalize it to all dimensional theory. We leave these tasks to future researches.

\section*{Acknowledgments}
This work is supported by the project funded by China Postdoctoral Science Foundation  with Grant No. 2021M691072, and the National Natural Science Foundation of China (NSFC) with Grants No. 12047519, No. 11875006 and No. 11961131013.

\appendix

\section{The curvature inherent in  $h^s_\alpha$}\label{app1}
To clarify the curvature inherent in the simplicity resolved holonomy $h^s_e$, let us consider its behaviour in continuum limit. Give a hyper-cubic graph $\gamma$ embedded in $\sigma$ with the coordinate length of the elementary edges of $\gamma$ being $\epsilon$. Then, we have the holonomy-flux phase space  $\times_{e\in E(\gamma)}T^\ast SO(D+1)_{e}$ associated to $\gamma$. One can proceed the gauge reduction with respect to the edge-simplicity constraint in this phase space, which leads to the reduced space composed by the elements $(h^s_{e},X_{e})_{e\in E(\gamma)}$, which are parametrized by twisted geometry parameters as
\begin{equation}
h^s_{e}=u_{e}e^{\xi^o_{e}\tau_o}\mathbb{I}^s\tilde{u}^{-1}_{e},\quad  X_{e}=\frac{1}{2}\eta_{e}V_{e},
\end{equation}
where $(\mathbb{I}^s)^{I}_{\ J}:=\delta_1^I\delta^1_J+\delta_2^I\delta^2_J$ is a $(D+1)\times (D+1)$ matrix. We can further solve the vertex simplicity constraint equation, and our following analysis is restricted on the constraint surface defined by vertex simplicity constraint in reduced space composed by $(h^s_{e},X_{e})_{e\in E(\gamma)}$.
  Notice that $h^s_e$ is a gauge (with respect to Gaussian constraint) covariant holonomy. To simplify our analysis, we can always proceed a gauge transformation to ensures
  \begin{equation}\label{gauge1}
 V^{IJ}_{e}=2\delta_1^{[I}v^{J]}_e, \quad \tilde{V}^{IJ}_{e}=2\delta_1^{[I}\tilde{v}^{J]}_e, \ \  \forall e\in E(\gamma).
  \end{equation}
Then, we have $(h_e^{\Gamma})^I_{\ J}\delta_1^J=\delta_1^I$ and
  \begin{equation}
 ( \Gamma_e)^I_{\ J}\delta_1^J=(\mathcal{O}(\epsilon^2))^I
  \end{equation}
with $\epsilon$ being small enough.
  To analysis the curvature captured by $h^s_e$, let us choose arbitrary minimal square loop $\alpha\subset \gamma$ composed by $\alpha=e_1\circ e_2\circ e_3\circ e_4$ and consider $h^s_{\alpha}=h^s_{e_1}h^s_{e_2}h^s_{e_3}h^s_{e_4}$. With the gauge conditions \eqref{gauge1} being satisfied, we can further fix a gauge which ensures
  \begin{equation}
  v^{I}_{e_1}=-v^{I}_{e_3},\ \    v^{I}_{e_2}=-v^{I}_{e_4},\ \   \text{and}\  \ v^{I}_{e_1}v^{J}_{e_2}\delta_{IJ}=0.
  \end{equation}
 By these conditions one have
  \begin{equation}
  (u_{e_1}\mathbb{I}^su^{-1}_{e_1}u_{e_2}\mathbb{I}^su^{-1}_{e_2})^I_{\ J}= (u_{e_2}\mathbb{I}^su^{-1}_{e_2}u_{e_3}\mathbb{I}^su^{-1}_{e_3})^I_{\ J}= (u_{e_3}\mathbb{I}^su^{-1}_{e_3}u_{e_4}\mathbb{I}^su^{-1}_{e_4})^I_{\ J}=\delta_1^I\delta^1_{J}.
  \end{equation}
  Then, recall
  \begin{equation}
  h_e^s=u_ee^{\xi\tau_o}\mathbb{I}^s\tilde{u}_e^{-1}\simeq ({u}_e\mathbb{I}^s{u}_e^{-1}+\beta K^{\perp}_e+\mathcal{O}(\epsilon^2))(\mathbb{I}+\Gamma_e+\mathcal{O}(\epsilon^2)),
  \end{equation}
   one can expand $h^s_{\alpha}$ as
\begin{eqnarray}
 (h^s_{\alpha})_{[KL]}\bar{\eta}_{I}^K\bar{\eta}_{J}^L&=&   (h^s_{e_1}h^s_{e_2}h^s_{e_3}h^s_{e_4})_{[KL]}\bar{\eta}_{I}^K\bar{\eta}_{J}^L\\\nonumber
   &=&\beta^2(K^{\perp}_{e_1}K^{\perp}_{e_4})_{[IJ]}+\mathcal{O}(\epsilon^3).
\end{eqnarray}
In continuum limit, it reads
\begin{eqnarray}\label{hsKK}
\lim_{\epsilon\to0}\frac{ (h^s_{\alpha})_{[KL]}\bar{\eta}_{I}^K\bar{\eta}_{J}^L }{\epsilon^2}
   &=&\beta^2\breve{K}_{a[I|K|}\breve{K}_{b\ \  J]}^{\ K}\dot{e}_1^a(v)\dot{e}_4^b(v)=-\beta^2{K}_{a[I}{K}_{b J]}\dot{e}_1^a(v)\dot{e}_4^b(v),
\end{eqnarray}
where $\breve{K}_{aIJ}:=K_{aIJ}-K_{aIJ}\bar{\eta}_{I}^K\bar{\eta}_{J}^L=2\delta^1_{[I}K^{}_{aJ]}$ with $\bar{\eta}_{I}^K:=\delta_I^K-\delta^1_I\delta_1^K$, $v$ is the source point of $e_1$ and target point of $e_4$, and $\dot{e}_1^a(v)$ and $\dot{e}_4^b(v)$ are the tangent vector of $e_1$ and $e_4$ at $v$ respectively. Thus, we conclude that $(h^s_{\alpha})_{[KL]}\bar{\eta}_{I}^K\bar{\eta}_{J}^L$ capture the degrees of freedom of extrinsic curvature properly.

\bibliographystyle{unsrt}

\bibliography{ref}

\end{document}